\documentstyle[11pt]{article}
\newcommand{\be}{\begin{eqnarray}}
\newcommand{\dlq}{\lq\lq}
\newcommand{\ee}{\end{eqnarray}}
\newcommand{\nhat}{\hat{n}}
\renewcommand{\baselinestretch}{1.0}
\title{
        \begin{flushright}
        {\normalsize
        NBI--98--21\\
        UALG/TP/98--6\\
        September 1998 \\}
        \end{flushright}
{\bf Non--perturbative computation of gluon mini--jet production in nuclear 
collisions at very high energies} 
       }
\author{Alex Krasnitz\\
       {\small\it UCEH, Universidade do Algarve 
        Campus de Gambelas, P-8000 Faro, Portugal}\\ 
        Raju Venugopalan\\ 
        {\small\it Niels Bohr Institute,
        Blegdamsvej 17,
        Copenhagen, Denmark, DK--2100 } \\          
       }

\date{}

\renewcommand{\baselinestretch}{1.5} 
\begin{document}
\setcounter{page}{0}
\maketitle
\thispagestyle{empty}
\begin{center}
{\bf Abstract}\\
\end{center}

\begin{small}
\noindent 
At very high energies, in the infinite momentum frame and in light cone
gauge, a hard scale proportional to the high parton density arises
in QCD. In an effective theory of QCD at small $x$,  this scale is of order
$\alpha_S\mu$, where $\mu$ is simply related to the gluon density at higher
rapidities. The {\it ab initio} real
time evolution of small $x$ modes in a nuclear collision can be described
consistently in the classical effective theory and various features of
interest can be studied non--perturbatively. In this paper, we discuss 
results from a real time SU(2) lattice computation of the production of 
gluon jets at very high
energies. At very large transverse momenta, $k_t\geq \mu$, our results match 
the predictions from pQCD based mini--jet calculations. 
Novel non--perturbative behaviour of the small $x$ modes is seen at smaller 
momenta $k_t\sim \alpha_S\mu$. Gauge invariant energy--energy correlators
are used to estimate energy distributions evolving in proper time.
\end{small}

\vfill \eject

\section{Introduction}
\vspace*{0.3cm}

It is of considerable theoretical and experimental interest to understand 
the collisions of nuclei at ultrarelativistic energies and the putative
evolution of the hot and dense matter created in these collisions into 
a thermalized, deconfined state of matter called a quark gluon plasma.
The theoretical challenge is to understand the dynamics of the formation of 
this matter and its properties from QCD while the experimental challenge is 
to detect evidence that such a plasma was indeed formed~\cite{QM96}.

The space--time evolution of the nuclei after the collision and the 
magnitudes and relevance of various proposed signatures of this hot and 
dense matter depend sensitively on the initial conditions for the evolution, 
namely, the parton distributions in each of the nuclei {\it prior} to the 
collision. In the standard perturbative QCD approach to
the problem, observables from the collision may be computed by convolving
the parton distributions of each nucleus, determined from deep 
inelastic scattering experiments, with the elementary parton--parton 
scattering cross sections. At the high energies of the RHIC and LHC 
colliders, hundreds of 
mini--jets may be formed in the initial collision~\cite{KajLanLin,EskKajLin,
BlaiMuell}. The final state interactions of these mini--jets are often
described in multiple scattering Glauber--Gribov models (see 
Ref.~\cite{Wang} and references therein) or in classical cascade 
approaches to obtain the space--time evolution (see Ref.~\cite{Geiger} 
and references therein).
The possible ``quenching'' of these mini--jets has also been studied 
and proposed as a signature of the formation of a quark gluon 
plasma~\cite{GyuPluWang}. 
Recently, initial conditions for the energy density and velocity obtained 
in the mini--jet approach have been used in a simple hydrodynamic model to 
study the late time evolution of matter in high energy nuclear
collisions~\cite{KKV,EskComment}.

While the above ``probabilistic'' approach provides a reasonable description
of large transverse momentum processes at large $x$, QCD coherence effects
become important as we go to small $x$ or alternatively, towards central
rapidities~\cite{comment}. This is because small $x$ partons in one nucleus
may ``see'' more than one parton in the direction of the incoming nucleus
resulting in a breakdown of the above described convolution of
distributions. What is needed therefore to describe the collision of the
``wee'' nuclear partons is a wave picture where coherent multiple scattering
is fully taken into account.  

At what values of $x$ would this picture become applicable? Strictly speaking,
coherence or shadowing effects start to become important when wee partons from
neighbouring nucleons overlap, i.e., when $x<1/(2R_N m_N)\approx 0.1$
fm. Here $R_N$ denotes the nucleon radius and $m_N$ the nucleon
mass. However, several authors have shown that with an appropriate choice of
the non--perturbative input structure functions, the standard leading twist
perturbative evolution equations still work fairly well in the shadowing
region below $x=0.1$\cite{Eskola,Kumano}.  To determine at precisely what
value of $x$ coherence effects become large one needs to compute the two
gluon distribution function at small $x$. Following the pioneering work of
Gribov, Levin and Ryskin~\cite{GLR} and of Mueller and Qiu~\cite{MuellQiu},
there has been a considerable body of recent work directed towards addressing
this issue~\cite{JKW,JKLW,MaQiuSterman}.

In this paper, we will describe an {\it ab initio} QCD based effective theory 
approach to the
theoretical study of nuclear collisions at very high energies. This model of
high energy nuclear collisions naturally incorporates coherence effects which
become important at small $x$ and small transverse momenta while reproducing
simultaneously the standard mini--jet results at large transverse momenta. It
has the further advantage of containing a self--consistent space--time
picture of the nuclear collison. The model is based on an effective action
approach to QCD initially developed by McLerran and 
Venugopalan~\cite{RajLar}, and later further developed by 
Ayala, Jalilian--Marian, McLerran and 
Venugopalan~\cite{AJMV}, and by J.Jalilian--Marian, Kovner, McLerran, Leonidov
and Weigert~\cite{JKMW,JKLW,JKW}.

The above mentioned effective action contains one dimensionful parameter,
$\chi(y,Q^2)$. (This parameter is also often used interchangeably with 
$\mu$ in the text;
the distinction between the two is discussed in section 3.) Here $\chi$ is
the total color charge squared per unit area integrated from the rapidity $y$
of interest to the beam rapidity. It is the only scale in the problem and we
expect therefore that the coupling constant runs as a function of this
scale. One therefore has weak coupling in the limits where the color charge
$\chi$ is large; either for $A\gg 1$ or $s\rightarrow \infty$.  It was argued
that the classical fields corresponding to the saddle point solutions 
of the effective theory are the non--Abelian
analogue of the Weizs\"acker--Williams fields in classical
electrodynamics. Exact analytical expressions for these fields have been
obtained recently~\cite{JKMW,Kovchegov}. Further, it has been shown
explicitly that $\chi$ obeys renormalization group equations in $y$ and $Q^2$.
These reduce to the well known BFKL and DGLAP equations
respectively~\cite{BFDGL} in the appropriate limits~\cite{JKLW,JKW}.

The above model was first applied to the problem of nuclear collisions by
Kovner, McLerran and Weigert, who formulated the problem as the collision of
Weizs\"acker--Williams fields~\cite{KLW}. The classical fields
after the collision then correspond to solutions of the Yang--Mills equations
in the presence of static, random sources of color charge on the light cone.
The initial conditions for the dynamical evolution of the small $x$ modes of
the two nuclei after the collision were formulated
and perturbative solutions obtained for modes with transverse momenta
$k_t>>\alpha_S\sqrt{\chi}$.  After averaging over the Gaussian random sources
of color charge on the light cone, 
the energy and number distributions of physical gluons were
computed. Further, the classical gluon radiation from these perturbative
modes was studied by these authors and later in greater detail by several
others~\cite{gyulassy,DirkYuri,SerBerDir}. In the small $x$ limit, it was shown
that the classical Yang--Mills result agreed with the quantum Bremsstrahlung 
result of Gunion and Bertsch~\cite{GunionBertsch}.

While the perturbative approach is very relevant and useful, it is still
essential to consider the full non--perturbative approach for the following
reasons. Firstly, the classical gluon radiation computed perturbatively is
infrared singular and has to be cut-off at some scale. This problem also
arises in mini--jet calculations where at high energies results are shown to
be rather sensitive to the cut--off~\cite{EskComment}. It was argued in
Ref.~\cite{KLW,DirkYuri} that a natural scale where the distributions are
cut-off is given by $k_t\sim \alpha_S \sqrt{\chi}$. However, since
quantitative differences can be large, it is important to perform a full
calculation. Secondly, the non--perturbative approach is crucial to study the
space--time evolution of the nuclei and in particular, the possible
thermalization of the system and the relevant time scales for
thermalization. This in turn has several ramifications for computations of
various signatures of the quark gluon plasma. For instance, if thermalization
does occur, then as proposed by Bjorken~\cite{Bj} hydrodynamic evolution of
the system is reasonable. In that event, our approach would provide the
initial temperature and velocity profiles necessary for such an
evolution~\cite{KKV} (see also~\cite{josef} and references therein).

We discuss in this paper results from real time simulations of the
full, non--perturbative evolution of classical non--Abelian
Weizs\"acker--Williams fields. Such a simulation is possible since the
fields are classical. Similar simulations of the real time evolution
of classical fields have been performed in the context of
sphaleron-mediated baryon number violation~\cite{Krasnitz} and
chirality violating transitions in hot gauge theories ~\cite{Moore}.

In brief, the idea is as follows~\cite{RajKrasnitz}.  We write down
the lattice Hamiltonian which describes the evolution of the small $x$
classical gauge fields.  It is the Kogut--Susskind Hamiltonian in
2+1--dimensions coupled to an adjoint scalar field. For simplicity, we
restrict our study in this paper to an SU(2) gauge theory. The
extension to the physical SU(3) case will be considered at a later
date. The lattice equations of motion for the fields are determined
straightforwardly using Hamilton's equations. The initial conditions
for the evolution are provided by the Weizs\"acker--Williams fields
for the nuclei before the collisions. Interestingly, the dependence on
the static light cone sources does not enter through the Hamiltonian
but instead from the initial conditions. Also, to reiterate, our
results have to be averaged over by the above mentioned Gaussian
measure for each source.

A limitation of our approach is that it is classical--quantum fluctuations
have been neglected. However if the effective action approach captures the
essential physics of the small $x$ modes of interest, then in the spirit of the
Wilson renormalization group, quantum information from the large x modes
(above the rapidity of interest) is contained in the parameter $\chi(y,Q^2)$
discussed above, which grows rapidly as one goes to smaller and smaller x's.
This information can be included in the classical lattice simulations. Thus
as long as we are not at small enough x where the above picture of Gaussian
random sources breaks down, the residual quantum fluctuations about the
classical saddle point of the effective theory should be small and the above
classical picture of nuclear collisions should be valid~\footnote{The 
classical picture of nuclear collisions will also be valid at very small $x$
but the Gaussian weight with which we average over the classical 
configurations will change. The functional replacing the Gaussian weight is
the solution of a non--linear renormalization group equation~\cite{JKW} and
is yet to be determined.}. 

A related approach is that of Mueller, Kovchegov and 
Wallon~\cite{MuellKovWall,MuellKov}, where they combined Mueller's dipole 
picture of high energy scattering~\cite{Mueller,MuellerPatel} with the 
classical Yang--Mills picture~\cite{Kovchegov} to study nucleon--nucleus 
scattering.  In particular, Mueller and Kovchegov make the interesting 
observation in their calculation of nucleon--nucleus scattering 
that, in light cone gauge, the effect of final state interactions is already 
contained in the wavefunction of the incoming hadron. This observation is
very much at the heart of this work because what final state interactions
there are, are very much determined by the initial conditions given by the
small $x$ wavefunctions of the nuclei before the collision.
For alternative approaches, we refer the reader to the work of Makhlin and 
Surdutovich~\cite{MakhSurd} and that of Balitskii~\cite{Ian}.

Our paper is organized as follows. In the following section we
discuss the problem of initial conditions for nuclear collisions as 
formulated by Kovner, McLerran
and Weigert and their perturbative solutions of the Yang--Mills equations and 
computation of gluon production in this
approach. We discuss a non--perturbative Hamiltonian approach to the 
solution of the full Yang--Mills equations.  

In section 3, we formulate the problem of solving the Yang--Mills
equations on the lattice. Assuming boost invariance and $N_c=2$, we
write down the lattice action in 2+1--dimensions. The lattice action
is used to construct the lattice analog of the continuum initial
conditions. This is done by matching the singular pieces of the
lattice equations of motion on the light cone. Once the initial
conditions are determined, we write down the lattice Hamiltonian and
the equations of motion for the evolution of the dynamical fields and
their conjugate momenta in the forward light cone.

In section 4, we use lattice perturbation theory to relate the parameters of
our lattice calculation, such as the parameter proportional to the parton
density, $\mu$, the lattice size $L$ and the lattice time $\tau$, to physical
strong interaction scales. We also discuss the relation of gauge invariant 
quantities
computed on the lattice to experimental observables that might be measured in
heavy ion collisions at RHIC and LHC.

Numerical results from our simulations are discussed in section 5.  These are
performed for a range of values of $g^2\mu$ = 0.018--0.2, and for lattice sizes
from $10\times 10$ to $160\times 160$, both measured in units of the lattice
spacing. (We will assume boost invariance throughout, which simplifies our
simulation to a 2--dimensional one.)  We first compare the field intensities
and the time evolution of hard modes in the collision with the predictions of
lattice perturbation theory and find excellent agreement. Significant
deviations from lattice perturbation theory are found for the softer modes 
$k_t \sim \alpha_S\mu$,
particularly at larger values of $\mu$. We study the dependence of our
results on the lattice size. We demonstrate the behaviour of the
chromo--electric and chromo--magnetic fields as a function of time and that 
of those components of the stress energy tensor that 
may be related to experimental observables.

We summarize our results in section 6 and discuss further computations that
may be performed in this approach. There are two appendices. The first
appendix discusses our numerical algorithm and procedure. The second is a
derivation of the lattice perturbation theory expressions for the field
intensity and the kinetic energy at $\tau=0$. These expressions were compared
to the full lattice results at large transverse momenta.

\section{Classical gluon radiation in high energy nuclear collisions}
\vspace*{0.3cm}

In the work of McLerran and Venugopalan~\cite{RajLar}, the classical gluon
field at small $x$ for a nucleus in the infinite frame is obtained by solving
the Yang--Mills equations in the presence of a static source of color charge
$\rho^a(r_t,\eta)$ on the light cone. This corresponds to the saddle point
solution of their small $x$ effective action. Exact solutions for the classical
field as functions of $\rho^a (r_t,\eta)$ were found by Jalilian-- Marian et
al.~\cite{JKMW} and independently by Kovchegov~\cite{Kovchegov}.
Distribution functions are computed by averaging products of the
classical fields over a Gaussian measure in $\rho$ with the variance
$\mu^2(\eta,Q^2)$. Here $\mu^2$ is the color charge squared per unit area per
unit rapidity resolved at a scale $Q^2$ by an external probe. It is related
to $\chi$ by the expression 
\be
\chi(\eta,Q^2)=\int_\eta^\infty d\eta^\prime
\mu^2(\eta^\prime,Q^2) \, .
\ee

The above picture of gluon fields in a nucleus at small $x$ was
extended to describe nuclear collisions by Kovner, McLerran and
Weigert~\cite{KLW}. 

In this section, we shall discuss their
formulation of the problem in the {\it continuum} and their
perturbative computation, to second order in the parameter
$\alpha_S\mu/k_t$, of classical gluon radiation in nuclear collisions. 
(Readers familiar with the discussion in Refs.~\cite{KLW}-\cite{SerBerDir} 
can skip sections 2.1 and 2.2 and go directly to 2.3.)
In section 2.3 we then briefly discuss a non--perturbative Hamiltonian approach
which suggests how all orders in $\alpha_S\mu/k_t$ can be computed
numerically.  The implementation of this approach on the lattice is
described in section 3.

\subsection{The non--Abelian Weizs\"acker--Williams approach to high energy 
nuclear collisions}
\vspace*{0.15cm}

In nuclear collisions at very high energies, the hard valence parton modes
act as highly Lorentz contracted, static sources of color charge for the
wee parton, Weizs\"acker--Williams modes in the nuclei. The sources are
described by the current
\be
J^{\nu,a}(r_t) = \delta^{\nu +}\rho_{1}^a (r_t)\delta(x^-) + \delta^{\nu -}
\rho_{2}^a (r_t) \delta(x^+) \, ,
\label{sources}
\ee
where $\rho_1$ and $\rho_2$ 
correspond to the color charge densities of the hard
modes in nucleus 1 and nucleus 2 respectively.  The classical field of two 
nuclei describing the wee parton dynamics is given by the solutions of 
the Yang--Mills
equations in the presence of two light cone sources.
We have then
\be
D_\mu F^{\mu\nu} = J^\nu \, .
\label{yangmill}
\ee

Gluon distributions are simply related to the Fourier transform $A_i^a (k_t)$ 
of the solution to the above equation by $<A_i^a(k_t) A_i^a(k_t)>_\rho$. The
averaging over the classical charge distributions is defined by
\be
\langle O\rangle_\rho &=& \int d\rho_{1}d\rho_{2}\, O(\rho_1,\rho_2) \nonumber \\
&\times& \exp\left( -\int d^2 r_t {{\rm Tr}\left[\rho_1^2(r_t)+\rho_2^2(r_t)
\right]
\over {2g^4\mu^2}}\right) \, .
\label{eqgauss}
\ee
The averaging over the color charge distributions is performed independently 
for each nucleus with equal Gaussian weight $g^4\mu^2$. (Note that this is
only true for the case of identical nuclei which will be assumed implicitly
throughout the rest of this paper.)

The observant reader will notice that we have omitted the rapidity dependence
of the the charge distributions in the equations immediately above. We will
justify this omission in our discussion of the lattice Hamiltonian. We note
that the rapidity dependence of the charge distribution is also absent in 
Ref.~\cite{KLW} (see the discussion below Eq.~\ref{yangmill2}).

Before the nuclei collide ($t<0$), a solution of the equations of motion is
\be
A^{\pm}&=&0 \, , \nonumber \\
A^i&=& \theta(x^-)\theta(-x^+)\alpha_1^i(r_t)+\theta(x^+)\theta(-x^-)
\alpha_2(r_t) \, ,
\label{befsoln}
\ee
where $\alpha_{q}^i(r_t)$ ($q=1,2$ denote the labels of the nuclei) 
are pure gauge fields defined through
the gauge transformation parameters $\Lambda_{q}(\eta,r_t)$ ~\cite{gyulassy}
\be
\alpha_{q}^i(r_t) = {1\over {i}}\left(Pe^{-i\int_{\pm \eta_{\rm proj}}^0
d\eta^\prime
\Lambda_{q}(\eta^\prime,r_t)}\right)
\nabla^i\left(Pe^{i\int_{\pm \eta_{\rm proj}}^0 d\eta^\prime 
\Lambda_{q}(\eta^\prime,r_t)}\right)^\dagger \, .
\label{puresoln}
\ee
Here $\eta=\eta_{\rm proj}-\log(x^-/x_{\rm proj}^-)$ is the rapidity of the 
nucleus 
moving along the positive light cone with the gluon field $\alpha_1^i$ and
$\eta=-\eta_{\rm proj}+\log(x_{\rm proj}^+/x^+)$ is the rapidity of the nucleus moving
along the negative light cone with the gluon field $\alpha_2^i$. 
The $\Lambda_{q}(\eta,r_t)$ parameters are, in turn, determined by the color
charge distributions:
\be
\Delta_\perp\Lambda_q=\rho_q \,\,;\,\, q=1,2 \,
\label{contpoi}\ee
$\Delta_\perp$ being the Laplacian in the perpendicular plane.

It is expected that at central rapidities (or $x\ll 1$) the source
density varies slowly as a function of rapidity and $\alpha^i\equiv
\alpha^i(r_t)$.  The above expression suggests that for $t<0$ the
solution is simply the sum of two disconnected pure gauges. Just as in
the Weizs\"acker--Williams limit in QED, the transverse components of
the electric field are highly singular.

For $t>0$ the solution is no longer pure gauge. Working in the Schwinger 
gauge
\be
x^+ A^- + x^- A^+ =0 \, ,
\label{schwinger}
\ee
or $A^\tau =0$, the  authors of Ref.~\cite{KLW} found that with the ansatz
\be
A^{\pm}&=&\pm x^{\pm}\alpha(\tau,r_t)\, , \nonumber \\
A^i&=&\alpha_\perp^i(\tau,r_t) \, ,
\label{ansatz}
\ee
where $\tau=\sqrt{2x^+ x^-}$, Eq.~\ref{yangmill} could be written in
the simpler form
\be
{1\over \tau^3}\partial_\tau \tau^3 \partial_\tau \alpha + [D_i,\left[D^i,
\alpha\right]]
&=&0 \, , \nonumber \\
{1\over \tau}[D_i,\partial_\tau \alpha_\perp^i] + i\tau[\alpha,\partial_
\tau \alpha] &=&0\, ,\nonumber \\
{1\over \tau}\partial_\tau \tau\partial_\tau \alpha_\perp^i
-i\tau^2[\alpha,\left[D^i,\alpha\right]]-[D^j,F^{ji}]&=&0 \, . 
\label{yangmill2}
\ee 
Note that the above equations of motion are independent of $\eta$--the 
gauge fields in the forward light cone are therefore only functions of
$\tau$ and $r_t$ and are explicitly boost invariant. We will use this 
fact later in our discussion of the Hamiltonian approach.

The initial conditions for the fields $\alpha(\tau,r_t)$ and $\alpha_\perp^i$
at $\tau =0$ are obtained by matching the equations of motion
(Eq.~\ref{yangmill}) at the point $x^\pm =0$ and along the boundaries
$x^+=0,x^->0$ and $x^-=0,x^+>0$. Because the sources are highly singular
functions along their respective light cones, so too in general will be 
the equations of
motion. Remarkably, however, there exists a set of non--singular initial
conditions that ensure the smooth evolution of the classical 
fields in the forward light
cone. These are obtained by matching the singular terms in the equations of 
motion before and after the collision at $\tau=0$. 
In terms of the fields of each of the nuclei
before the collision ($t<0$), the fields after the collision ($t>0$) are
\be
\alpha_\perp^i|_{\tau=0}&=& \alpha_1^i+\alpha_2^i \, , \nonumber \\
\alpha|_{\tau=0}&=&{i\over 2} [\alpha_1^i,\alpha_2^i] \, .
\label{initial}
\ee
Gyulassy and McLerran have shown~\cite{gyulassy} that even when the fields
$\alpha_{1,2}^i$ before the collision are smeared out in rapidity to 
properly account for singular contact terms in the equations of motion the
above boundary conditions remain unchanged.
Further, since the equations are very singular at $\tau=0$, the
only condition on the derivatives of the fields that would lead to regular
solutions are $\partial_\tau \alpha|_{\tau=0},\partial_\tau \alpha_\perp^i
|_{\tau=0} =0$.

\subsection{Review of perturbative solution of the Yang--Mills equations}
\vspace*{0.15cm}

In Ref.~\cite{KLW}, perturbative solutions (for small $\rho$) were found 
to order $\rho^2$ by expanding 
the initial conditions and the fields in powers $\rho$ (or equivalently,
in powers of $\alpha_S\mu/k_t$) as  
\be
\alpha = \sum_{n=0}^{\infty} \alpha_{(n)}\,\, ; \,\, \alpha_{\perp}^i 
= \sum_{n=0}^{\infty} \alpha_{\perp (n)}^i \, ,
\ee
where the subscript $n$ denotes the $n$th order in $\rho$. Since it is 
relevant to the discussion in the following sections (particularly appendix B)
, we outline their
solution below with a commentary but refer the reader to their paper for the 
details.

To lowest order in $\rho$, the Yang--Mills equations of motion are
linear in $\alpha_{(1)}$ and $\alpha_{\perp (1)}^i$. The solution is
the trivial Abelian solution
\be
\alpha_{(1)}=0\,\,;\,\, \alpha_{\perp (1)}^i = -\partial^i (\phi^1 +
\phi^2)(x_t)\, ,
\ee
where $\phi^q = {-1\over \nabla_{\perp}^2}\rho^q$. Clearly $\alpha_{\perp
(1)}$ above is a pure gauge. To second order in $\rho$, the equations of 
motion are
nearly homogeneous except for a non--linear term proportional to
$[\alpha_{\perp (1)}^j, \alpha_{\perp (1)}^i]$ in the equation for
$\alpha_{\perp (2)}^i$. This piece can be removed from the equations of
motion to this order by making use of the residual, $\tau$--independent gauge
transformation $A_\mu = V(x_t)[\epsilon_\mu -{1\over i}\partial_\mu]\
V^\dagger (x_t)$ and fixing $V$ such that $\epsilon^i$ satisfies the two
dimensional Coulomb gauge condition $\partial^i \epsilon^i|_{\tau=0}=0$.
As described in appendix B, the  lattice Coulomb gauge condition is 
fixed in an analogous way to eliminate the residual gauge freedom on the 
lattice.

Writing $\epsilon^i=\epsilon^{ij}\partial^j \chi$, where $\epsilon^{ij}$ is
the two dimensional anti--symmetric Levi--Civita tensor, the Yang--Mills
equations in the forward light cone (Eq.~\ref{yangmill2}) may be written as
\be
{1\over \tau^3}\partial_\tau \tau^3 \partial_\tau
\epsilon_{(2)}-\nabla_{\perp}^2 \epsilon_{(2)} &=& 0 \, , \nonumber  \\
{1\over \tau} \partial_\tau \tau \partial_\tau \chi_{(2)} -\nabla_{\perp}^2
\chi_{(2)} &=& 0 \, . 
\label{ym2ndord}
\ee
with the initial conditions
\be
\epsilon_{(2)}|_{\tau=0} &=& {i\over 2} 
[\partial^i \phi^1,\partial^i \phi^2] \, ,\nonumber \\
\chi_{(2)}|_{\tau=0} &=& -i\epsilon^{ij} [\partial^i \phi^1,\partial^j \phi^2]
\, .
\ee
The residual Coulomb gauge fixing at this second order therefore shifts the 
non--linearities from the equations of motion to the initial conditions.

The solutions of Eqs.~\ref{ym2ndord} can be written in terms of Bessel
functions as
\be
\epsilon_{(2)}(\tau,x_t) &=& \int {d^2 k_t d^2 y_t\over {(2\pi)^2}}
e^{ik_t\cdot (x-y)_t} h_3(y_t) {1\over {\omega\tau}} J_1 (\omega\tau) \, ,
\nonumber \\
\chi_{(2)}(\tau,x_t) &=& \int {d^2 k_t d^2 y_t\over {(2\pi)^2}}
e^{ik_t\cdot (x-y)_t} h_1(y_t) J_0 (\omega\tau) \, ,
\label{pertsols}
\ee
where 
\be
h_3 (y_t) &=& i [\partial^i\phi^1,\partial^i\phi^2](y_t)\, ,\nonumber \\
h_1 (y_t) &=& -i \epsilon^{ij}{1\over \nabla_{\perp}^2}[\partial^i\phi^1,
\partial^j\phi^2] (y_t) \, .
\label{h1h3}
\ee
Here $\omega = |k_t|$. 
Note that in the solutions to the gauge fields above, the spatial and 
temporal distributions factorize. The amplitudes of the fields in this 
perturbative approach are therefore completely determined at $\tau=0$.
In section 5, we will see that in weak coupling the large transverse momentum
modes show the above Bessel behavior. (In lattice perturbation theory, the 
form is the same as above. The difference is that $\omega$ obeys a 
lattice dispersion relation).

At late times $\omega\tau >>1$, the well known asymptotic expressions for
the Bessel functions can be used to write the solutions in Eq.~\ref{pertsols}
as 
\be
\epsilon_{(2)}(\tau,x_t) &=& \int {d^2 k_t\over {(2\pi)^2}}
{1\over\sqrt{2\omega}}\left\{ a_1(\vec{k_t}){1\over \tau^{3/2}} 
e^{ik_t\cdot x_t-i\omega\tau} + h.c\right\} \, , \nonumber \\
\epsilon^i (\tau,x_t) &=& \int {d^2 k_t\over {(2\pi)^2}}\kappa^i
{1\over\sqrt{2\omega}}\left\{ a_2(\vec{k_t}){1\over \tau^{1/2}} 
e^{ik_t\cdot x_t-i\omega\tau} + h.c\right\} \, .
\ee
Here $\kappa^i = \epsilon^{ij}k_t^j/\omega$ and 
\be
a_1(k_t) = {1\over \sqrt{\pi}} {h_3(k_t)\over \omega}\,\,;\,\, 
a_2 (k_t)={1\over \sqrt{\pi}}{i\omega h_1(k_t)} \, ,
\ee
where $h_1$ and $h_3$ are now the Fourier transforms of Eq.~\ref{h1h3}.

The energy distribution in a transverse box of size $R$ 
and longitudinal extent $dz$ can be computed by summing over the energy of 
the modes in the box with the occupation number of the modes given by
the functions $a_i(k_t)$ above. We have then (for $\omega\tau>>1$) 
\be
{dE\over {dy d^2k_t}} = {1\over {(2\pi)^2}}\sum_{i,b} |a_i^b(k_t)|^2 \, .
\ee
The multiplicity distribution of classical gluons is 
defined as $dE/dyd^2 k_t/\omega$. After performing the averaging over the
Gaussian sources, the number distribution of classical gluons is
\be
{dN\over {dyd^2 k_t}} = \pi R^2 {2g^6 \mu^4\over {(2\pi)^4}} {N_c (N_c^2-1)
\over k_t^4} L(k_t,\lambda) \, ,
\label{GunBer}
\ee
where $L(k_t,\lambda)$ is an infrared divergent function at the scale
$\lambda$. It will be discussed further below. We first note that this result
agrees with the quantum bremsstrahlung formula of Gunion and
Bertsch~\cite{GunionBertsch} and with several later
works~\cite{gyulassy,DirkYuri,SerBerDir}.  It was also shown by Gyulassy and
McLerran that when the sources are smeared in rapidity, the expression that
results is identical to the one above except $\mu^4\rightarrow
\chi^+(y)\chi^-(y)$ where the $\pm$ superscripts refer to the nucleus on the
positive or negative light cone respectively.

The origin of the infrared divergent function $L(k_t,\lambda)$ above is 
from long range color correlations which are cut-off either by a nuclear 
form factor (as in Refs.~\cite{GunionBertsch,DirkYuri}), by dynamical screening
effects~\cite{GyuWang,EskMullWang} or in the classical Yang--Mills case 
of Ref.~\cite{KLW}, non--linearities that become large at the scale 
$k_t\sim \alpha_S\mu$. In the classical case then,
\be
L(k_t,\lambda) = \log(k_t^2/\lambda^2)\, ,
\ee
where $\lambda = \alpha_S\mu$. A similar logarithmic behaviour at
small transverse momenta can be deduced from the dipole form factors used in
Refs.~\cite{GunionBertsch,GyuWang}. A power law ($\sim 1/k_t^2$) infrared behaviour
is predicted in Ref.~\cite{DirkYuri}. The formalism used in all these 
derivations
breaks down at small momenta and one cannot distinguish between the different
parametrizations of the nuclear form factors. However, at sufficiently high
energies, the behaviour of $L(k_t,\lambda)$ in the infrared is given by
higher order (in $\alpha_S\mu/k_t$) non--linear terms in the classical
effective theory. One of the goals of our work is to address precisely this 
question: how do non--perturbative effects in the classical 
effective theory change the
gluon distributions at small transverse momenta?

\subsection{The Hamiltonian approach}
\vspace*{0.3cm}

While the Yang--Mills equations can be solved perturbatively, in the
limit $\alpha_S \mu \ll k_t$, it is unlikely that a simple analytical
solution exists for Eq.~\ref{yangmill} in the non--perturbative regime 
where $k_t\leq \mu$.
The classical solutions have to be determined numerically for
$t>0$. The straightforward procedure would be to discretize
Eq.~\ref{yangmill}.  It will be more convenient for our purposes
though to construct the lattice Hamiltonian and obtain the lattice
equations of motion from Hamilton's equations. This will be done in
the next section. Before we do that, we will discuss here the form of
the continuum Hamiltonian and comment on our assumption of boost
invariance.

We start from the QCD action (without dynamical quarks) 
\be
S_{QCD} = \int d^4 x \sqrt{-g}\left\{{1\over 4} g^{\mu\lambda}g^{\nu\sigma}
F_{\mu\nu}F_{\lambda\sigma}-j^\mu A_\mu \right\} \, ,
\ee
where $g={\rm det}(g_{\mu\nu})$. In the forward light cone ($t>0$) it is 
convenient to work with the $\tau,\eta,\vec{r_t}$ co--ordinates where 
$\tau=\sqrt{2 x^+ x^-}$ is the proper 
time, $\eta={1\over 2}\log(x^+/x^-)$ is the 
space--time rapidity and $\vec{r_t}=(x,y)$ are the two transverse Euclidean 
co--ordinates. In these co--ordinates, the metric is diagonal with 
$g^{\tau\tau}=-g^{xx}=-g^{yy}=1$ and $g^{\eta\eta}=-1/\tau^2$.

After a little algebra, the Hamiltonian can be written as~\cite{Sasha}
\be
H =\tau\int d\eta {\rm d}^2r_t\left\{{1\over 2} p^{\eta,a}p^{\eta,a}
+{1\over {2\tau^2}}p^{r,a} p^{r,a} + {1\over{2\tau^2}} F_{\eta r}^aF_{\eta r}^a
+{1\over 4}F_{ij}^aF_{ij}^a + j^{\eta,a} A_\eta^a + j^{r,a} A_r^a\right\} \, .
\label{hamilton}
\ee
Here we have adopted the gauge condition of Eq.~\ref{schwinger}, which is 
equivalent to requiring $A^\tau =0$. 
Also, $p^\eta={1\over \tau}\partial_\tau A_\eta$ and $p^r=\tau \partial_\tau 
A_r$ are the conjugate momenta.

Consider the field strength $F_{\eta r}$ in the above Hamiltonian. If we 
assume approximate boost invariance, or
\be 
A_r (\tau,\eta,\vec{r_t})\approx A_r(\tau,\vec{r_t}); \ \  
A_{\eta}(\tau,\eta,\vec{r_t})\approx \Phi(\tau,\vec{r_t}),
\label{adef}
\ee
we obtain  
\be
F_{\eta r}^a = -D_r \Phi^a \, ,
\label{fdstrgth}
\ee
where $D_r =\partial_r -ig A_r$ is the covariant derivative. Further, if we 
express $j^{\eta,r}$ in terms of the $j^{\pm}$ defined in 
Eq.~\ref{sources} we obtain the result that $j^{\eta,r}=0$ for $\tau>0$.

Performing the integration over the space--time rapidity, we can re--write
the Hamiltonian in Eq.~\ref{hamilton} as 
\be
H = \int d\vec{r_t} \eta \left\{ {1\over 2\tau} E_r^a E_r^a +
{\tau\over 4}F_{ij}^a F_{ij}^a + {1\over {2\tau}} \left(D_r
\Phi\right)^a \left(D_r \Phi\right)^a + {\tau\over 2}p_\eta^a
p_\eta^a\right\} \, .
\label{twodh}
\ee
We have thus succeeded in expressing
the Hamiltonian in Eq.~\ref{hamilton} as the Yang--Mills Hamiltonian in
2+1--dimensions coupled to an adjoint scalar. 
The discrete version of the
above Hamiltonian is well known and is the Kogut--Susskind
Hamiltonian~\cite{KS} in 2+1--dimensions coupled to an adjoint scalar field.
The lattice Hamiltonian will be discussed further in the next section.

We now comment on a key assumption in the above derivation,
namely, the boost invariance of the fields.  This invariance results in
Eq.~\ref{twodh} thereby allowing us to restrict ourselves to a transverse
lattice alone. To clarify the issue we are compelled to make a few historical
remarks. As we mentioned earlier, the authors of Ref.~\cite{KLW} found a
solution which was explicitly boost invariant.  However, this result was a
consequence of the original assumption of McLerran and Venugopalan that the
color charge density factorizes, $\rho^a(r_t,\eta)\rightarrow \rho^a(r_t)
\delta(x^-)$. It was noticed in Ref.~\cite{RajGavai} that this factorized
form for the charge density results in infrared singular correlation
functions which diverge as the square of the lattice size.  This problem was
subsequently resolved in Ref.~\cite{JKMW} where the authors realized that a
rapidity dependent charge density $\rho^a(r_t,\eta)$ would give infrared safe
solutions. This might be interpreted as implying that the boost invariance
assumption of Ref.~\cite{KLW} should be given up as well.

Fortunately, this is not necessary. In principle, the rapidity dependence of
the color charge density can be arbitrarily weak since that is sufficient to
obtain infrared safe correlation functions. In Ref.~\cite{RajLar2}, an
explicit model was constructed for the color charge distribution in the
fragmentation region.  It was shown there that for $\eta<\eta_{\rm proj}$ the
color charge distribution had a very weak dependence on $\eta$. Further, as
mentioned earlier, it was shown by Gyulassy and
McLerran~\cite{gyulassy} that the initial conditions in Eq.~\ref{initial} are
unaffected by the smearing in rapidity.

In general, at the energies of interest, particle distributions are unlikely
to be boost invariant. Eskola, Kajantie and Ruuskanen~\cite{KKV} have shown
in the mini--jet picture that the final distributions are more like broad
Gaussians. This would be true in our case as well since $\mu^2$ is in general
a function of the rapidity and this dependence may be strong in the central
region.  If we wish to describe particle distributions for a range of
rapidities as opposed to our current restriction to 1 unit in rapidity, we
will have to give up our assumption of boost invariance. This can be easily
done, but it will be numerically more time consuming as well.

\section{Real-time lattice description of nuclear collisions}
\vspace*{0.3cm}

In the previous section we discussed the continuum formulation of nuclear 
collisions in terms of collisions of non--Abelian Weizs\"acker--Williams 
fields. The initial conditions for the evolution after the collision were 
obtained by requiring that the singular contributions to the equations of
motion on the light cone vanish. We next discussed the perturbative solutions
of the equations of motion for our choice of initial conditions and obtained 
a result for the number distribution of radiated gluons. 

In this section, 
we will formulate the problem on the lattice. We begin by writing down the
lattice action and equations of motion. From the lattice action, we can 
indentify the singular terms in the lattice equations of motion. Matching 
these on the light cone uniquely gives us the initial
conditions on the lattice. We end with a discussion of Hamilton's equations 
of motion for the evolution of dynamical fields on the lattice with the 
initial conditions specified at $\tau =0$.

\subsection{Lattice action and equations of motion}
\vspace*{0.15cm}
In this sub--section, we will write down the lattice action in
2+1--dimensions and the rules to derive the lattice equations of
motion from this action. In the following sub--section, we will
identify the singular terms in the lattice action and use these to
derive the correct initial conditions for dynamical evolution of
fields on the lattice for $\tau>0$.

The action is defined in the 4-space discretized in the transverse directions,
while $z$ and $t$ are continuous. The appropriate expression is derived
starting from the Minkowski Wilson action in the discretized 4-space and
taking the naive continuum limit in the longitudinal directions. The Minkowski
Wilson action for the $SU(N_c)$ gauge group in fundamental representation reads
\begin{eqnarray}
S&=&a^{-2}\sum_{zt}\left(1-{1\over N_c}\Re\,{\rm Tr}U_{zt}\right)
+\sum_{t\perp}\left(1-{1\over N_c}\Re\,{\rm Tr}U_{t\perp}\right) \nonumber \\
&&-\sum_{z\perp}\left(1-{1\over N_c}\Re\,{\rm Tr}U_{z\perp}\right)
-a^2\sum_{\perp}\left(1-{1\over N_c}\Re\,{\rm Tr}U_{\perp}\right),\nonumber
\end{eqnarray}
where $zt$, $z\perp$, $t\perp$ and $\perp$ are, in obvious notation, plaquettes
lying in various 2-planes of the 4-space and $\Re$ denotes the real component.
A plaquette is defined as 
\be
U_{jlm}\equiv U_{j,l}U_{j+l,m}U^\dagger_{j+m,l}U^\dagger_{j,m}\, .\nonumber
\ee
where $j$ is a site index and $l,m$ are direction indices. We now take the 
formal continuum limit in the longitudinal directions by writing longitudinal
links as $U=\exp(ia A)$,
letting $a\rightarrow 0$ and assuming that all the fields are smooth functions
of longitudinal coordinates. The powers of $a$ in front of various terms in $S$
have been chosen with this formal limit in mind. Replacing $a^2\sum_{zt}$ with
$\int{\rm d}z{\rm d}t$, we then have for the action
\begin{equation}
S=\int{\rm d}z{\rm d}t\sum_\perp\left[{1\over {2N_c}}{\rm Tr}F_{zt}^2
+{1\over N_c}\Re\,{\rm Tr}(M_{t\perp}-M_{z\perp})
-\left(1-{1\over N_c}\Re\,{\rm Tr}U_{\perp}\right)\right],\label{action}
\label{twodact}
\end{equation}
where
\begin{equation}
M_{t,jn}\equiv {1\over 2}(A_{t,j}^2+A_{t,j+n}^2)-U_{j,n}\left[{1\over 2}
\partial_t^2U^\dagger_{j,n}+
i(A_{t,j+n}\partial_tU^\dagger_{j,n}-\partial_tU^\dagger_{j,n}A_{t,j})
+A_{t,j+n}U^\dagger_{j,n}A_{t,j}\right]\, ,
\label{mterm}
\end{equation}
and similarly for $M_{z,jn}$. 

The equation of motion for a field is obtained by varying $S$ with respect to
that field. For the longitudinal fields $A_{t,z}$ the variation has the usual
meaning of a partial derivative. For transverse link matrices $U_\perp$ the 
variation amounts to a covariant derivative
\be
D^\gamma S(U_\perp)\equiv \partial_r
S\left(\exp(ir\sigma_\gamma)U_\perp\right)|_{r=0} \, ,
\label{rule}
\ee
where $\sigma_\gamma$, $1\leq \gamma\leq N_c^2-1$ form a 
basis of $SU(N_c)$ algebra..
In particular,
\begin{equation}D^\gamma U_\perp=i\sigma_\gamma U_\perp; \ \ 
D^\gamma U_\perp^\dagger=-iU_\perp^\dagger\sigma_\gamma,\label{du}
\end{equation}
and derivatives of more complicated functions are obtained, combining these two
rules with the usual rules of differentiation.

\subsection{Initial conditions on the lattice}
\vspace*{0.15cm}

We now derive the lattice analogue of the continuum initial conditions in
Eq.~\ref{initial}. We start from the lattice action in Eq.~\ref{twodact},
obtain the lattice equations of motion in the four light cone regions and
determine non--singular initial conditions by matching at $\tau=0$ the
coefficients of the most singular terms in the equations of motion.

On the lattice, the initial conditions are the constraints on the
longitudinal gauge potential $A^\pm$ and the transverse link matrices
$U_\perp$ at $\tau=0$.  
The longitudinal gauge potentials are zero outside the light cone
and satisfy the Schwinger gauge condition (Eq.~\ref{schwinger}) inside the
light cone $x_\pm>0$. Thus they can be written as in the continuum case (see
Eq.~\ref{ansatz}) as
\begin{equation}
A^\pm=\pm x^\pm\theta(x^+)\theta(x^-)\alpha(\tau, x_t)\, .
\label{apm}
\end{equation}
From our discussion in the last section, it is clear that the transverse link
matrices are, for each nucleus, pure gauges before the
collision. This fact is reflected by writing
\begin{equation}
U_\perp=\theta(-x^+)\theta(-x^-)I+\theta(x^+)\theta(x^-)U(\tau)
+\theta(-x^+)\theta(x^-)U^{(1)}+\theta(x^+)\theta(-x^-)U^{(2)} \, ,
\label{uperp}
\end{equation}
where $U^{(1),(2)}$ are pure gauge. 

The pure gauges are defined on the lattice as follows. To each lattice
site $j$ we assign two SU($N_c$) matrices $V_{1,j}$ and
$V_{2,j}$. Each of these two defines a pure gauge lattice gauge
configuration with the link variables
\be
U_{j,\nhat}^{(q)} = V_{q,j}V_{q,j+n}^\dagger \, ,
\label{pgs}
\ee
where $q=1,2$ labels the two nuclei. As in the continuum, the gauge 
transformation matrices $V_{q,j}$ are determined by the color charge 
distribution $\rho_{q,j}$ of the nuclei, normally distributed with the 
standard deviation $g^4\mu_L^2$ (compare with the continuum distribution in 
Eq.~\ref{eqgauss}): 
\be
P[\rho_q]\propto\exp\left(-{1\over{2g^4\mu_L^2}}\sum_j\rho_{q,j}^2\right).
\label{distriblat}\ee
Parametrizing $V_{q,j}$ as $\exp(i\Lambda^q_j)$ with Hermitean traceless
$\Lambda^q_j$, we then obtain $\Lambda^q_j$ by solving the lattice
Poisson equation
\be
\Delta_L\Lambda^q_j\equiv\sum_n\left(\Lambda^q_{j+n}+\Lambda^q_{j-n}
-2\Lambda^q_j\right)=\rho_{q,j}.
\label{latpoi}\ee
It is easy to verify that the correct 
continuum solution (Eqs.~\ref{befsoln} and \ref{ansatz}) for the transverse 
fields $A_\perp$ is recovered by taking the formal continuum limit of 
Eq.~\ref{uperp}. Using the general representation of the gauge fields in
Eqs.~\ref{apm} and~\ref{uperp}, we shall now derive the initial conditions
for them at $\tau=0$.

\subsection*{Initial conditions for $x^\pm>0$: $U_\perp$}

The equation of motion for $U_\perp$, contains, upon substitution of
$U_\perp$ from (\ref{uperp}) and $A^\pm$ from (\ref{apm}), two types of terms.
\begin{enumerate}
\item Terms regular at $x^\pm=0$. All the terms containing less
than two longitudinal derivatives belong to this category. The regularity of 
terms containing no longitudinal derivatives is clear. Terms in the action
containing a single longitudinal derivative all involve the combination
$A^-\partial_-+A^+\partial_+$. Given the functional form 
(\ref{apm}) of $A^\pm$, these terms only give rise to regular contributions to
the equation of motion. The double-derivative contributions 
$\Re{\rm Tr}U_\perp^\dagger\partial_+\partial_-U_\perp$ in the action
give rise to terms in the equation of motion behaving as 
\be
{1\over\tau}x^\pm\delta(x^\pm)\theta(x^\mp)\partial_\tau U \, , \nonumber
\ee
which is regular if $\partial_\tau U$ vanishes rapidly enough as 
$\tau\rightarrow 0$. The terms regular at $x_\pm=0$ provide no relation 
between $U_\perp$ and $U^{(1),(2)}$.
\item The singular terms containing the product $\delta(x^+)\delta(x^-)$.
These originate in the double-derivative contributions 
$\Re{\rm Tr}U_\perp^\dagger\partial_+\partial_-U_\perp$ in the action, when
both derivative operators act on the step functions. Since the coefficient
in front of $\delta(x^+)\delta(x^-)$ must vanish in order to satisfy the
equation of motion, a matching relation between $U$ and $U^{(1),(2)}$ is 
obtained.
\end{enumerate}

We now compute the singular part of the $U_\perp$ equation of motion. We note
that 
\be
U_\perp(\partial_t^2-\partial_z^2)U^\dagger_\perp
=2U_\perp(\partial_+\partial_-U^\dagger_\perp)=
2(\partial_+\partial_-U_\perp)U^\dagger_\perp \nonumber
\ee
modulo a total derivative and utilize
the identity $\theta(x)\delta(x)={1\over 2}\delta(x)$ with the rules
of Eq.~\ref{du}. This gives, upon substitution of (Eq.~\ref{uperp})
and retaining only the terms proportional to $\delta(x^+)\delta(x^-)$,
\begin{equation}
{\rm Tr}\,\sigma_\gamma\left[(U^{(1)}+U^{(2)})(I+U^\dagger)-
{\rm h.c.}\right]=0\, .
\label{ucond}
\end{equation}
We therefore have obtained the result that $(U^{(1)}+U^{(2)})
(I+U^\dagger)$ should
have no anti-Hermitean traceless part. Note that this condition has the
correct formal continuum limit: writing $U^{(1),(2)}$ as
$\exp(ia_\perp\alpha_{1,2})$ and $U$ as $\exp(ia_\perp\alpha_\perp)$, we
have, for small $a_\perp$,
$$\alpha_\perp=\alpha_1+\alpha_2,$$ as required.

The above condition in Eq.~(\ref{ucond}) can easily be resolved in the SU(2) 
case, because
the sum of any two $2\times 2$ unitary matrices is a unitary matrix times a
real number. Also, the anti-Hermitean part of an SU(2) matrix is traceless, 
while the Hermitean part is proportional to I. Therefore in this case 
Eq.~\ref{ucond} is equivalent to 
\be
(U^{(1)}+U^{(2)})(I+U^\dagger)=CI \, ,\nonumber
\ee
where $C$ is a real number. Resolving this equation for $U^\dagger$ and 
requiring that $U^\dagger U=I$, we obtain, after a simple calculation, that
$C={\rm Tr}(U^{(1)}+U^{(2)})$ and hence
\be
U={{{\rm Tr}(U^{(1)}+U^{(2)})}\over{{U^{(1)}}^\dagger+{U^{(2)}}^\dagger}}-I
=(U^{(1)}+U^{(2)})({U^{(1)}}^\dagger+{U^{(2)}}^\dagger)^{-1} \, .
\label{ucondp}
\ee
For $N_c>2$ the solution can be obtained by solving the $N_c^2-1$
equations (\ref{ucond}) for the $N_c^2-1$ real numbers parametrizing $U$.
The solution does not necessarily have the compact form (\ref{ucondp}).
For simplicity, we leave the $N_c>2$ case outside the scope of this paper.

\subsection*{Initial conditions for $x^\pm>0$: $A^\pm$}

We have seen that the matching conditions for the fields before and after the
collision stem from the requirement that the singular terms in the equations
of motion cancel out. We now impose this requirement on the equation of motion
for $A_-$. As before, the singularities arise only because longitudinal
derivatives of step functions are taken.

Consider first the $F_{+-}^2$ term in the action. Since in this case we need
two derivatives for a genuine singularity, 
we are only interested in the Abelian part of $F_{+-}^2$, whose
variation with respect to $A^{+,\gamma}$ gives
$${1\over N_c}{\rm Tr}\sigma_\gamma\partial_+(\partial_-A_+-\partial_+A_-),$$
whose most singular part is (using Eq.~\ref{apm} and $x\delta^\prime (x)
=-\delta(x)$)
\be
\alpha_\gamma\theta(x^-)\delta(x^+)\, .\nonumber 
\ee
We now vary the $\pm,\perp$ terms (Eq.~\ref{mterm}) in the action
(Eq.~\ref{action}) with respect to $A_j^{+,\gamma}$ and select the
contributions containing derivatives. The result is
\be
-{i\over{2N_c}}\sum_n{\rm Tr}\sigma_\gamma
\left[(\partial_+U^\dagger_{j-n,n})U_{j-n,n}
-U^\dagger_{j-n,n}(\partial_+U_{j-n,n})-U_{j,n}(\partial_+U^\dagger_{j,n})
+(\partial_+U_{j,n})U^\dagger_{j,n}\right] \, ,\nonumber
\ee
the singular part of which is
\be
&-{i\over{2N_c}}\delta(x^+)\sum_n{\rm Tr}\sigma_\gamma
&\Big[\theta(x^-)(U^{(2)}-{U^{(2)}}^\dagger+U^\dagger U^{(1)}-
{U^{(1)}}^\dagger U)_{j-n,n} 
\nonumber \\
&&-\theta(x^-)(U^{(2)}-{U^{(2)}}^\dagger+U^{(1)} U^\dagger-U {U^{(1)}}
^\dagger)_{j,n} \nonumber \\
&&+({U^{(2)}}^\dagger-U^{(2)})_{j-n,n}-({U^{(2)}}^\dagger-U^{(2)})_{j,n}\Big]
 \, . 
\label{dmda}
\ee
Requiring that the coefficient in front of $\delta(x^+)\theta(x^-)$ vanish, we
obtain
\be
\alpha_\gamma={i\over{2N_c}}\sum_n{\rm Tr}\sigma_\gamma
\left[(U^{(2)}+U^{(1)}U^\dagger-{\rm h.c.})_{j,n}
-(U^{(2)}+U^\dagger U^{(1)}-{\rm h.c.})_{j-n,n}\right]\, . \nonumber
\ee
This result can be cast in a more symmetric form, using the initial condition
for $U_\perp$. Note that
\be
U^{(2)}+U^{(1)}U^\dagger={1\over 2}(U^{(1)}+U^{(2)})(I+U^\dagger)
+{1\over 2}(U^{(1)}-U^{(2)})(U^\dagger-I)\, . \nonumber
\ee
Since the first term on the right-hand side has no anti-Hermitean traceless 
part, it follows that
\be
\alpha_\gamma={i\over{4N_c}}\sum_n{\rm Tr}\sigma_\gamma
\left([(U^{(1)}-U^{(2)})(U^\dagger-I)-{\rm h.c.}]_{j,n}
-[(U^\dagger-I)(U^{(1)}-U^{(2)})-{\rm h.c.}]_{j-n,n}\right)\, .\nonumber \\
\label{agamma}
\ee
It is easily seen that the above equation has the correct formal continuum
limit.  Writing again $U^{(1),(2)}$ as $\exp(ia_\perp\alpha_{1,2})$ and $U$ as
$\exp(ia_\perp\alpha_\perp)$, we have, for small $a_\perp$,
\be
(U^{(1)}-U^{(2)})(U^\dagger-I)-{\rm h.c.}\approx 2[\alpha_1,\alpha_2]\, .\nonumber
\ee
Hence, in the limit of smooth fields,
\be
\alpha=i\sum_n[\alpha_1,\alpha_2]_n\, ,\nonumber
\ee
as required.

Finally, we require that the last line of Eq.~\ref{dmda} be cancelled by the
term in the action coupling $A^+$ to the external source. This term then must
have the form
\be
{i\over{2N_c}}\int{\rm d}z{\rm d}t\sum_j\delta(x^+){\rm Tr}A_j^+\sum_n
\left[({U^{(2)}}^\dagger-U^{(2)})_{j-n,n}-({U^{(2)}}^\dagger-U^{(2)})_{j,n}\right]\, ,\nonumber 
\ee
whose formal continuum limit is easily seen to be the correct one:
\be
-{1\over N_c}\int{\rm d}^4x\delta(x^+){\rm Tr}A^+\nabla\cdot A_\perp \, .
\nonumber 
\ee
A completely analogous term should be included for $A^-$.

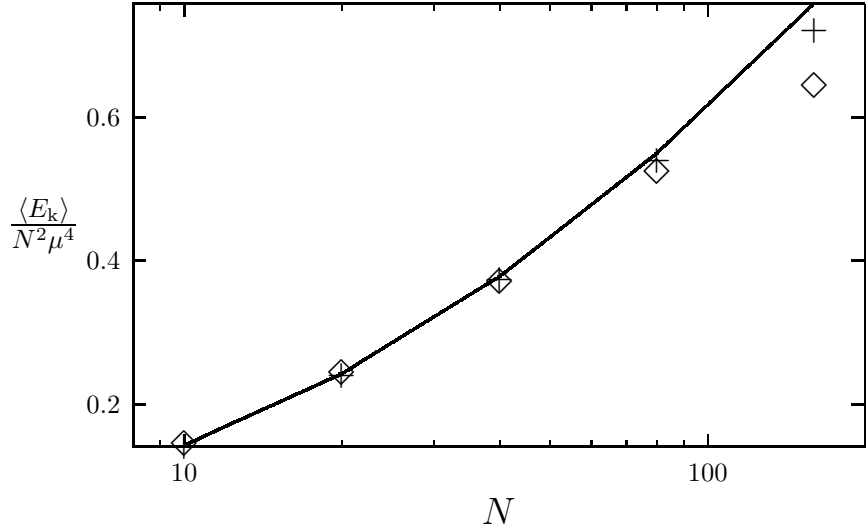
\begin{figure}[ht]
\setlength{\unitlength}{0.240900pt}
\ifx\plotpoint\undefined\newsavebox{\plotpoint}\fi
\sbox{\plotpoint}{\rule[-0.200pt]{0.400pt}{0.400pt}}%
\begin{picture}(1350,900)(0,0)
\begin{Large}
\font\gnuplot=cmr10 at 10pt
\gnuplot
\sbox{\plotpoint}{\rule[-0.200pt]{0.400pt}{0.400pt}}%
\put(181.0,229.0){\rule[-0.200pt]{4.818pt}{0.400pt}}
\put(161,229){\makebox(0,0)[r]{0.2}}
\put(1310.0,229.0){\rule[-0.200pt]{4.818pt}{0.400pt}}
\put(181.0,455.0){\rule[-0.200pt]{4.818pt}{0.400pt}}
\put(161,455){\makebox(0,0)[r]{0.4}}
\put(1310.0,455.0){\rule[-0.200pt]{4.818pt}{0.400pt}}
\put(181.0,680.0){\rule[-0.200pt]{4.818pt}{0.400pt}}
\put(161,680){\makebox(0,0)[r]{0.6}}
\put(1310.0,680.0){\rule[-0.200pt]{4.818pt}{0.400pt}}
\put(181.0,163.0){\rule[-0.200pt]{0.400pt}{2.409pt}}
\put(181.0,849.0){\rule[-0.200pt]{0.400pt}{2.409pt}}
\put(223.0,163.0){\rule[-0.200pt]{0.400pt}{2.409pt}}
\put(223.0,849.0){\rule[-0.200pt]{0.400pt}{2.409pt}}
\put(261.0,163.0){\rule[-0.200pt]{0.400pt}{4.818pt}}
\put(261,122){\makebox(0,0){10}}
\put(261.0,839.0){\rule[-0.200pt]{0.400pt}{4.818pt}}
\put(508.0,163.0){\rule[-0.200pt]{0.400pt}{2.409pt}}
\put(508.0,849.0){\rule[-0.200pt]{0.400pt}{2.409pt}}
\put(653.0,163.0){\rule[-0.200pt]{0.400pt}{2.409pt}}
\put(653.0,849.0){\rule[-0.200pt]{0.400pt}{2.409pt}}
\put(756.0,163.0){\rule[-0.200pt]{0.400pt}{2.409pt}}
\put(756.0,849.0){\rule[-0.200pt]{0.400pt}{2.409pt}}
\put(835.0,163.0){\rule[-0.200pt]{0.400pt}{2.409pt}}
\put(835.0,849.0){\rule[-0.200pt]{0.400pt}{2.409pt}}
\put(900.0,163.0){\rule[-0.200pt]{0.400pt}{2.409pt}}
\put(900.0,849.0){\rule[-0.200pt]{0.400pt}{2.409pt}}
\put(955.0,163.0){\rule[-0.200pt]{0.400pt}{2.409pt}}
\put(955.0,849.0){\rule[-0.200pt]{0.400pt}{2.409pt}}
\put(1003.0,163.0){\rule[-0.200pt]{0.400pt}{2.409pt}}
\put(1003.0,849.0){\rule[-0.200pt]{0.400pt}{2.409pt}}
\put(1045.0,163.0){\rule[-0.200pt]{0.400pt}{2.409pt}}
\put(1045.0,849.0){\rule[-0.200pt]{0.400pt}{2.409pt}}
\put(1083.0,163.0){\rule[-0.200pt]{0.400pt}{4.818pt}}
\put(1083,122){\makebox(0,0){100}}
\put(1083.0,839.0){\rule[-0.200pt]{0.400pt}{4.818pt}}
\put(1330.0,163.0){\rule[-0.200pt]{0.400pt}{2.409pt}}
\put(1330.0,849.0){\rule[-0.200pt]{0.400pt}{2.409pt}}
\put(181.0,163.0){\rule[-0.200pt]{276.794pt}{0.400pt}}
\put(1330.0,163.0){\rule[-0.200pt]{0.400pt}{167.666pt}}
\put(181.0,859.0){\rule[-0.200pt]{276.794pt}{0.400pt}}
\put(41,511){\makebox(0,0){${{\langle E_{\rm k}\rangle}\over{N^2\mu^4}}$}}
\put(755,61){\makebox(0,0){$N$}}
\put(181.0,163.0){\rule[-0.200pt]{0.400pt}{167.666pt}}
\put(261,165){\raisebox{-.8pt}{\makebox(0,0){$\Diamond$}}}
\put(508,277){\raisebox{-.8pt}{\makebox(0,0){$\Diamond$}}}
\put(756,419){\raisebox{-.8pt}{\makebox(0,0){$\Diamond$}}}
\put(1003,592){\raisebox{-.8pt}{\makebox(0,0){$\Diamond$}}}
\put(1250,727){\raisebox{-.8pt}{\makebox(0,0){$\Diamond$}}}
\sbox{\plotpoint}{\rule[-0.400pt]{0.800pt}{0.800pt}}%
\put(261,165){\usebox{\plotpoint}}
\multiput(261.00,166.41)(1.106,0.501){217}{\rule{1.964pt}{0.121pt}}
\multiput(261.00,163.34)(242.923,112.000){2}{\rule{0.982pt}{0.800pt}}
\multiput(508.00,278.41)(0.811,0.501){299}{\rule{1.497pt}{0.121pt}}
\multiput(508.00,275.34)(244.893,153.000){2}{\rule{0.748pt}{0.800pt}}
\multiput(756.00,431.41)(0.637,0.500){381}{\rule{1.219pt}{0.121pt}}
\multiput(756.00,428.34)(244.471,194.000){2}{\rule{0.609pt}{0.800pt}}
\multiput(1003.00,625.41)(0.525,0.500){463}{\rule{1.041pt}{0.121pt}}
\multiput(1003.00,622.34)(244.840,235.000){2}{\rule{0.520pt}{0.800pt}}
\sbox{\plotpoint}{\rule[-0.500pt]{1.000pt}{1.000pt}}%
\sbox{\plotpoint}{\rule[-0.600pt]{1.200pt}{1.200pt}}%
\put(261,163){\makebox(0,0){$+$}}
\put(508,274){\makebox(0,0){$+$}}
\put(756,426){\makebox(0,0){$+$}}
\put(1003,612){\makebox(0,0){$+$}}
\put(1250,817){\makebox(0,0){$+$}}
\end{Large}
\end{picture}
\vskip-.6cm
\caption{The lattice size dependence of the scalar kinetic energy density,
expressed in units of $\mu^4$ for $\mu=0.0177$ (pluses) and $\mu=0.035$
(diamonds). The solid line is the LPTh prediction. The error bars are smaller
than the plotting symbols.}
\label{ekvsl}
\end{figure}

\subsection{Hamiltonian formulation on the lattice}
\vspace*{0.15cm}

In the previous sub-sections, we wrote down the lattice action and
lattice equations of motion for all regions of the light
cone. Matching the singular terms on the light cone in section 3.2, we
obtained the initial conditions for evolution in the forward light
cone.

In section 2.3, we derived the continuum Hamiltonian for the forward 
light cone ($\tau >0$) to be the
Yang--Mills Hamiltonian in 2+1--dimensions in the gauge $A^\tau=0$.
The lattice Hamiltonian is obtained by performing a Legendre transform
of Eq.~\ref{twodact} following the standard Kogut-Susskind
procedure~\cite{KS}.  The analog of the Kogut--Susskind Hamiltonian
here is
\be
H_L&=& {1\over{2\tau}}\sum_{l\equiv (j,\nhat)} 
E_l^{a} E_l^{a} + \tau\sum_{\Box} \left(1-
{1\over 2}{\rm Tr} U_{\Box}\right)  \, ,\nonumber \\
&+& {1\over{4\tau}}\sum_{j,\nhat}{\rm Tr}\,
\left(\Phi_j-U_{j,\nhat}\Phi_{j+\nhat}
U_{j,\nhat}^\dagger\right)^2 +{\tau\over 4}\sum_j {\rm Tr}\,p_j^2,
\label{hl}\ee
where $E_l$ are generators of right covariant derivatives on the group
and $U_{j,\nhat}$ is a component of the usual SU(2) matrices corresponding
to a link from the site $j$ in the direction $\nhat$. The first two terms
correspond to the contributions to the Hamiltonian from the chromoelectric and
chromomagnetic field strengths respectively. In the last equation, 
$\Phi\equiv \Phi^a\sigma^a$ is the adjoint scalar field with its conjugate
momentum $p\equiv p^a\sigma^a$. Taking the continuum limit of the above 
Hamiltonian, one recovers the continuum Hamiltonian in Eq. \ref{twodh}.

Lattice equations of motion follow directly from $H_L$ of Eq.~\ref{hl}.  For
any dynamical variable $v$ with no explicit time dependence ${\dot
v}=\{H_L,v\}$, where ${\dot v}$ is the derivative with respect to $\tau$, and
$\{\}$ denote Poisson brackets. We take $E_l$, $U_l$, $p_j$, and $\Phi_j$ as
independent dynamical variables, whose only nonvanishing Poisson brackets are
$$\{p_i^a,\Phi_j^b\}=\delta_{ij}\delta_{ab}; \ \ 
\{E_l^a,U_m\}=-i\delta_{lm}U_l\sigma^a; \ \
\{E_l^a,E_m^b\}=2\delta_{lm}\epsilon_{abc}E_l^c$$
(no summing of repeated indices). The equations of motion are consistent with
a set of local constraints (Gauss' laws). These are
\begin{equation}
C^a_j\equiv \sum_{\nhat}
\left[{1\over 2}E_{j,{\nhat}}^b{\rm Tr}\left(\sigma^a U_{j,\nhat}\sigma^b
U_{j,{\nhat}}^\dagger\right)
-E_{j-{\nhat},{\nhat}}^a\right]
-2\epsilon_{abc}p^b_j\Phi^c_j=0.
\label{calpha}\end{equation}

Explicitly, using the Poisson brackets above, the equations of motion
for the four dynamical variables are as follows:
\be
{\dot U}_m &=& {-i\over \tau} U_m E_l \, ,\\
{\dot E}_l &=& \left\{\tau\sum_{\Box} \left(1-
{1\over 2}{\rm Tr} U_{\Box}\right), E_l\right\} \nonumber \\
&-& {i\over \tau} \left({\tilde \Phi}_j \Phi_{j+\nhat}-\Phi_{j+\nhat}{\tilde 
\Phi}_j\right) \, , \\
{\dot \Phi}_j &=& \tau p_j \, \\
{\dot p}_j &=& {1\over \tau} \left[ {\tilde \Phi}_{j+\nhat}^\dagger
+{\tilde \Phi}_{j-\nhat}-2\Phi_j\right] \, .
\label{lateqs}
\ee
Above, ${\tilde \Phi} = U_{j,\nhat} \Phi U_{j,\nhat}^\dagger$.

The results of section 3 can be summarized as follows. The four
independent dynamical variables are $E_l$, $U_\perp$, $p_j$ and $\Phi_j$. 
Their evolution in $\tau$ after the nuclear collision is determined by
Hamilton's equations above and their values at the initial 
time $\tau =0$ are specified by the initial conditions derived explicitly in 
section 3.2. For easy reference, these are written compactly as follows:
\be
U |_{\tau=0} &=& (U_1+U_2)(U_1^\dagger + U_2^\dagger)^{-1} \,\, ; 
\,\, E_l |_{\tau=0} = 0 \, . \nonumber \\
p_j |_{\tau=0} &=& 2\alpha \,\, ; \,\, \Phi_j = 0 \, ,
\label{dinitial}
\ee
where $U$ and $\alpha$ are given by Eq.~\ref{ucondp} and
Eq.~\ref{agamma} respectively. Note that the second set of conditions for 
$\Phi_j$ and $p_j$ follows respectively from Eq.~\ref{adef} and the
definition of $p_j$--see the discussion after Eq.~\ref{hamilton}.

\section{Interpreting lattice results for continuum physics}
\vspace*{0.3cm}

Before we turn to discussing in detail the numerical results from our lattice
simulations, we should consider first the ramifications for the problem of
nuclear collisions at very high energies. Specifically, we wish to know how 
one interprets in physical terms the results of our numerical simulations. 

In section 2, we introduced a scale $\mu^2$, the color
charge squared per unit area, and argued that new non--perturbative physics
arises at momenta of order $k_t\sim \alpha_S\mu$. In the original work of
McLerran and Venugopalan, only the valence quarks were taken to be sources of
color charge which gave $\mu^2\sim A^{1/3}$ fm$^{-2}$. Hence $\mu \gg
\Lambda_{QCD}$ only for nuclei much larger than physical nuclei. However, if
semi--hard gluons at x values greater than those of interest 
here~\footnote{Note that in this
case, these correspond to those values of x corresponding to rapidities
greater than the central rapidity in the nuclear collision.} are also included
as sources of color charge as they should be, then $\mu^2$ is significantly 
larger. Then $\mu^2$ is defined as~\cite{gyulassy}
\be
\mu^2 = {A^{1/3}\over {\pi r_0^2}} \int_{x_0}^1 dx \left({1\over 2 N_c} 
q(x,Q^2) + {N_c\over {N_c^2-1}} g(x,Q^2)\right) \, ,
\ee
where $q,g$ stand for the {\it nucleon} 
quark and gluon structure functions at the resolution 
scale $Q$ of the physical process of interest. Also, above $x_0 =
Q/\sqrt{s}$. Using the HERA structure function data, Gyulassy and McLerran
estimated that $\mu\leq 1$ GeV for LHC energies and $\mu \leq 0.5$ GeV 
at RHIC. Thus the regime where the classical Yang--Mills picture can be
applied is rather limited. However, 
a window does exist and depending on what higher order calculations will tell
us, this window may be larger or smaller than the naive classical extimates. 
Moreover, the insights gained from this self--consistent spacetime
approach may still be very valuable in modelling heavy ion collisions which
are not at the asymptotic energies where the approach becomes quantitative.

This work is not a quantitative study to be compared to experiment. Rather, 
our goal is a qualitative understanding of non--perturbative phenomena in
the central region of a heavy ion collision. Towards this end, we need to 
\begin{itemize}
\item
relate lattice parameters to physical scales,
\item 
define and compute quantities on the lattice that
may be eventually compared to experiments with as little ambiguity as possible.
\end{itemize}

Let us first relate lattice units and physical units. Given $A$ of a
nucleus, we should match valence parton densities. To this end,
we require the equality of cross-sectional areas. The
relation between the linear size $L=Na$ of the lattice
and the transverse radius $R$ of the nucleus is then $L^2=\pi R^2$.
We ignore for the moment the difference in the lattice and continuum 
boundary conditions. For an $A=200$ nucleus $R=6.55$ fm translates into
$L=11.6$ fm. We can then express the lattice spacing $a=L/N$ in units
of fermi.

\begin{figure}[ht]
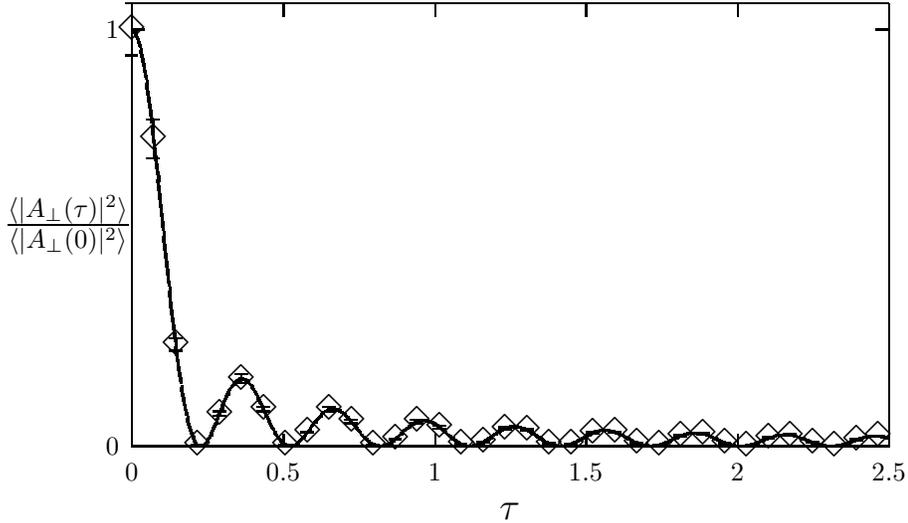

\setlength{\unitlength}{0.240900pt}
\ifx\plotpoint\undefined\newsavebox{\plotpoint}\fi
\sbox{\plotpoint}{\rule[-0.200pt]{0.400pt}{0.400pt}}%

\vskip-0.6cm
\caption{Normalized field intensity of a hard ($k_t=2.16{\rm GeV}$) mode 
vs proper time $\tau$ in units of fm (diamonds). Solid line is the LPTh 
prediction.}
\label{a2kvspht025pil160mu025}
\end{figure}

Next, we determine the relation between the continuum $\mu$ and its
lattice counterpart. To do so, we use the Poisson equation
(\ref{latpoi}) for the gauge transformation parameter $\Lambda$:
$\Delta_L \Lambda=\rho_L$, and compare it to the continuum equation
(\ref{contpoi}): $\Delta_\perp\Lambda=\rho_c$. For $a\rightarrow 0$
the lattice Laplacian $\Delta_L$ approaches $a^2\Delta$, hence we must
have $\rho_L=\rho_ca^2$. Also, as $a\rightarrow 0$, the lattice and
continuum distributions of the color charge should become
identical. Keeping in mind that $a^2\sum=\int {\rm d}x_\perp^2$ for
small $a$, we obtain $\mu_L=a\mu_c$. In the effective theory on the
lattice the coupling is $g^2 \mu_L L$~\cite{RajGavai}.  In this
qualitative work, for simplicity, we choose $g=1$.  Then $\mu_L =
0.1$, for instance, for an $A=200$ nucleus corresponds to
\be
\mu_c = 
{0.1\over a} \rightarrow 0.1 \,{N\over L} \sim 0.28\,\,\mbox{GeV for
N=160} \, .
\ee
For a physical scale of say $\mu=0.6$ GeV, and the number of sites
$N=160$, we should pick $\mu_L\sim 0.21$.  To compare lattice momenta
to momenta in physical units, note that $k_t = 2\pi n/L$, where
$-(N-1)/2 \leq n \leq (N-1)/2$. Then, for $L=11.6$ fm, $k_t\sim 1$ GeV
for $n\sim $ 9--10. For $\mu_c=1$ GeV or equivalently, $\mu_L = 0.36$,
one may expect perturbation theory to be reliable for $n\geq 10$. One
can in principle estimate the value of $n$ where strong coupling
effects become significant.

There still remains the question of how to relate lattice time
$\tau_{\rm latt}$ to the time in physical units, $\tau_{\rm
phys}$. This can be obtained by comparing the continuum and lattice
Hamiltonians at small a,~\cite{Krasnitz}
\be
\tau_{\rm phys} = a \tau_{\rm lattice} \nonumber \, .
\ee
Recall that $a=L/N$ fm. For $L= 11.6$ and $N=160$, $a=0.07$ fm.

Now that we have straightened out the relation of lattice units to physical
units, we must consider what can be measured on the lattice. Interesting 
quantities would be those for which our simulations would predict non--
perturbative corrections to ``empirical'' quantities which can be computed
in perturbative QCD at large momenta. A quantity one may consider is 
the cross section for gluon mini--jet production. At large transverse momenta,
Gyulassy and McLerran~\cite{gyulassy} have shown that the classical 
Yang--Mills formula in Eq.~\ref{GunBer} is at small $x$ (approximately) 
the same as the perturbative QCD prediction~\cite{GLR} for the process 
$AA\rightarrow g$ 
\be
{d\sigma\over {dyd^2 k_t}} = K_N {\alpha_S N_c\over {\pi^2 k_t^2}}
\int d^2 q_t {f(x_1,q_t^2) f(x_2,(\vec{k}_t-\vec{q}_t)^2)\over {q_t^2 
(\vec{k}_t-\vec{q}_t)^2}} \, ,
\ee
where 
\be
f(x,Q^2) = {d xG(x,Q^2) \over {d\log{Q^2}}} \, ,
\ee
and $x_1\approx x_2 = k_t/\sqrt{s}$. The two formulae are equivalent if we
divide the above formula by $\pi R^2$, approximate the integral above by
factoring out $f$ above at the scale $k_t^2$ and taking the normalization
factor $K_N\approx 5$. At large transverse momenta therefore, we can relate
the field intensity measured on the lattice $|A_{k_t}|^2$ for the appropriate
values of $\mu$ to the cross section for $AA\rightarrow g$. The lattice
perturbation theory expression (which matches the prediction of the full
theory at large momenta) is given by Eq.~\ref{dcrfin} of appendix B.

However, at smaller transverse momenta, there is no simple relation between
the field intensity and the classical gluon distribution function (and
thereby the cross section by the above arguments). This is because the
dispersion relation for soft gluons is not that for free gluons--the 
interpretation of the field intensity as the gluon distribution is then
clearly not right. We have to look for more
general gauge invariant 
quantities which, conversely, in the limit of large $k_t$, will give
us the $AA\rightarrow g$ cross section.

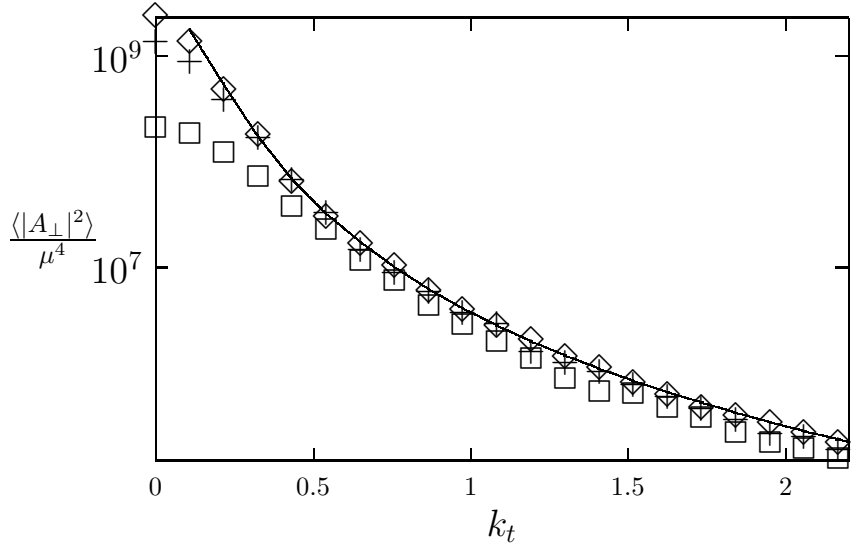
\begin{figure}[ht]
\setlength{\unitlength}{0.240900pt}
\ifx\plotpoint\undefined\newsavebox{\plotpoint}\fi
\sbox{\plotpoint}{\rule[-0.200pt]{0.400pt}{0.400pt}}%
\begin{picture}(1350,900)(0,0)
\begin{Large}
\font\gnuplot=cmr10 at 10pt
\gnuplot
\sbox{\plotpoint}{\rule[-0.200pt]{0.400pt}{0.400pt}}%
\put(241.0,466.0){\rule[-0.200pt]{4.818pt}{0.400pt}}
\put(221,466){\makebox(0,0)[r]{$10^7$}}
\put(1310.0,466.0){\rule[-0.200pt]{4.818pt}{0.400pt}}
\put(241.0,798.0){\rule[-0.200pt]{4.818pt}{0.400pt}}
\put(221,798){\makebox(0,0)[r]{$10^9$}}
\put(1310.0,798.0){\rule[-0.200pt]{4.818pt}{0.400pt}}
\put(241.0,163.0){\rule[-0.200pt]{0.400pt}{4.818pt}}
\put(241,122){\makebox(0,0){0}}
\put(241.0,839.0){\rule[-0.200pt]{0.400pt}{4.818pt}}
\put(489.0,163.0){\rule[-0.200pt]{0.400pt}{4.818pt}}
\put(489,122){\makebox(0,0){0.5}}
\put(489.0,839.0){\rule[-0.200pt]{0.400pt}{4.818pt}}
\put(736.0,163.0){\rule[-0.200pt]{0.400pt}{4.818pt}}
\put(736,122){\makebox(0,0){1}}
\put(736.0,839.0){\rule[-0.200pt]{0.400pt}{4.818pt}}
\put(983.0,163.0){\rule[-0.200pt]{0.400pt}{4.818pt}}
\put(983,122){\makebox(0,0){1.5}}
\put(983.0,839.0){\rule[-0.200pt]{0.400pt}{4.818pt}}
\put(1231.0,163.0){\rule[-0.200pt]{0.400pt}{4.818pt}}
\put(1231,122){\makebox(0,0){2}}
\put(1231.0,839.0){\rule[-0.200pt]{0.400pt}{4.818pt}}
\put(241.0,163.0){\rule[-0.200pt]{262.340pt}{0.400pt}}
\put(1330.0,163.0){\rule[-0.200pt]{0.400pt}{167.666pt}}
\put(241.0,859.0){\rule[-0.200pt]{262.340pt}{0.400pt}}
\put(80,511){\makebox(0,0){${{\langle |A_\perp|^2\rangle}\over{\mu^4}}$}}
\put(785,61){\makebox(0,0){$k_t$}}
\put(241.0,163.0){\rule[-0.200pt]{0.400pt}{167.666pt}}
\put(241,859){\raisebox{-.8pt}{\makebox(0,0){$\Diamond$}}}
\put(295,818){\raisebox{-.8pt}{\makebox(0,0){$\Diamond$}}}
\put(348,743){\raisebox{-.8pt}{\makebox(0,0){$\Diamond$}}}
\put(402,672){\raisebox{-.8pt}{\makebox(0,0){$\Diamond$}}}
\put(455,599){\raisebox{-.8pt}{\makebox(0,0){$\Diamond$}}}
\put(509,544){\raisebox{-.8pt}{\makebox(0,0){$\Diamond$}}}
\put(563,501){\raisebox{-.8pt}{\makebox(0,0){$\Diamond$}}}
\put(616,466){\raisebox{-.8pt}{\makebox(0,0){$\Diamond$}}}
\put(670,428){\raisebox{-.8pt}{\makebox(0,0){$\Diamond$}}}
\put(723,398){\raisebox{-.8pt}{\makebox(0,0){$\Diamond$}}}
\put(777,373){\raisebox{-.8pt}{\makebox(0,0){$\Diamond$}}}
\put(831,351){\raisebox{-.8pt}{\makebox(0,0){$\Diamond$}}}
\put(884,323){\raisebox{-.8pt}{\makebox(0,0){$\Diamond$}}}
\put(938,306){\raisebox{-.8pt}{\makebox(0,0){$\Diamond$}}}
\put(991,282){\raisebox{-.8pt}{\makebox(0,0){$\Diamond$}}}
\put(1045,264){\raisebox{-.8pt}{\makebox(0,0){$\Diamond$}}}
\put(1098,243){\raisebox{-.8pt}{\makebox(0,0){$\Diamond$}}}
\put(1152,231){\raisebox{-.8pt}{\makebox(0,0){$\Diamond$}}}
\put(1206,220){\raisebox{-.8pt}{\makebox(0,0){$\Diamond$}}}
\put(1259,204){\raisebox{-.8pt}{\makebox(0,0){$\Diamond$}}}
\put(1313,188){\raisebox{-.8pt}{\makebox(0,0){$\Diamond$}}}
\sbox{\plotpoint}{\rule[-0.400pt]{0.800pt}{0.800pt}}%
\put(241,822){\makebox(0,0){$+$}}
\put(295,790){\makebox(0,0){$+$}}
\put(348,731){\makebox(0,0){$+$}}
\put(402,670){\makebox(0,0){$+$}}
\put(455,604){\makebox(0,0){$+$}}
\put(509,553){\makebox(0,0){$+$}}
\put(563,494){\makebox(0,0){$+$}}
\put(616,459){\makebox(0,0){$+$}}
\put(670,428){\makebox(0,0){$+$}}
\put(723,396){\makebox(0,0){$+$}}
\put(777,378){\makebox(0,0){$+$}}
\put(831,335){\makebox(0,0){$+$}}
\put(884,317){\makebox(0,0){$+$}}
\put(938,303){\makebox(0,0){$+$}}
\put(991,283){\makebox(0,0){$+$}}
\put(1045,264){\makebox(0,0){$+$}}
\put(1098,244){\makebox(0,0){$+$}}
\put(1152,227){\makebox(0,0){$+$}}
\put(1206,205){\makebox(0,0){$+$}}
\put(1259,200){\makebox(0,0){$+$}}
\put(1313,181){\makebox(0,0){$+$}}
\sbox{\plotpoint}{\rule[-0.500pt]{1.000pt}{1.000pt}}%
\sbox{\plotpoint}{\rule[-0.600pt]{1.200pt}{1.200pt}}%
\put(241,684){\raisebox{-.8pt}{\makebox(0,0){$\Box$}}}
\put(295,674){\raisebox{-.8pt}{\makebox(0,0){$\Box$}}}
\put(348,645){\raisebox{-.8pt}{\makebox(0,0){$\Box$}}}
\put(402,607){\raisebox{-.8pt}{\makebox(0,0){$\Box$}}}
\put(455,560){\raisebox{-.8pt}{\makebox(0,0){$\Box$}}}
\put(509,524){\raisebox{-.8pt}{\makebox(0,0){$\Box$}}}
\put(563,474){\raisebox{-.8pt}{\makebox(0,0){$\Box$}}}
\put(616,443){\raisebox{-.8pt}{\makebox(0,0){$\Box$}}}
\put(670,404){\raisebox{-.8pt}{\makebox(0,0){$\Box$}}}
\put(723,374){\raisebox{-.8pt}{\makebox(0,0){$\Box$}}}
\put(777,347){\raisebox{-.8pt}{\makebox(0,0){$\Box$}}}
\put(831,320){\raisebox{-.8pt}{\makebox(0,0){$\Box$}}}
\put(884,289){\raisebox{-.8pt}{\makebox(0,0){$\Box$}}}
\put(938,268){\raisebox{-.8pt}{\makebox(0,0){$\Box$}}}
\put(991,265){\raisebox{-.8pt}{\makebox(0,0){$\Box$}}}
\put(1045,243){\raisebox{-.8pt}{\makebox(0,0){$\Box$}}}
\put(1098,228){\raisebox{-.8pt}{\makebox(0,0){$\Box$}}}
\put(1152,205){\raisebox{-.8pt}{\makebox(0,0){$\Box$}}}
\put(1206,188){\raisebox{-.8pt}{\makebox(0,0){$\Box$}}}
\put(1259,179){\raisebox{-.8pt}{\makebox(0,0){$\Box$}}}
\put(1313,163){\raisebox{-.8pt}{\makebox(0,0){$\Box$}}}
\sbox{\plotpoint}{\rule[-0.500pt]{1.000pt}{1.000pt}}%
\sbox{\plotpoint}{\rule[-0.200pt]{0.400pt}{0.400pt}}%
\put(295,840){\usebox{\plotpoint}}
\multiput(295.58,836.92)(0.498,-0.803){103}{\rule{0.120pt}{0.742pt}}
\multiput(294.17,838.46)(53.000,-83.461){2}{\rule{0.400pt}{0.371pt}}
\multiput(348.58,752.06)(0.498,-0.760){105}{\rule{0.120pt}{0.707pt}}
\multiput(347.17,753.53)(54.000,-80.532){2}{\rule{0.400pt}{0.354pt}}
\multiput(402.58,670.49)(0.498,-0.632){103}{\rule{0.120pt}{0.606pt}}
\multiput(401.17,671.74)(53.000,-65.743){2}{\rule{0.400pt}{0.303pt}}
\multiput(455.58,603.89)(0.498,-0.509){105}{\rule{0.120pt}{0.507pt}}
\multiput(454.17,604.95)(54.000,-53.947){2}{\rule{0.400pt}{0.254pt}}
\multiput(509.00,549.92)(0.600,-0.498){87}{\rule{0.580pt}{0.120pt}}
\multiput(509.00,550.17)(52.796,-45.000){2}{\rule{0.290pt}{0.400pt}}
\multiput(563.00,504.92)(0.680,-0.498){75}{\rule{0.644pt}{0.120pt}}
\multiput(563.00,505.17)(51.664,-39.000){2}{\rule{0.322pt}{0.400pt}}
\multiput(616.00,465.92)(0.773,-0.498){67}{\rule{0.717pt}{0.120pt}}
\multiput(616.00,466.17)(52.512,-35.000){2}{\rule{0.359pt}{0.400pt}}
\multiput(670.00,430.92)(0.887,-0.497){57}{\rule{0.807pt}{0.120pt}}
\multiput(670.00,431.17)(51.326,-30.000){2}{\rule{0.403pt}{0.400pt}}
\multiput(723.00,400.92)(1.005,-0.497){51}{\rule{0.900pt}{0.120pt}}
\multiput(723.00,401.17)(52.132,-27.000){2}{\rule{0.450pt}{0.400pt}}
\multiput(777.00,373.92)(1.087,-0.497){47}{\rule{0.964pt}{0.120pt}}
\multiput(777.00,374.17)(51.999,-25.000){2}{\rule{0.482pt}{0.400pt}}
\multiput(831.00,348.92)(1.215,-0.496){41}{\rule{1.064pt}{0.120pt}}
\multiput(831.00,349.17)(50.792,-22.000){2}{\rule{0.532pt}{0.400pt}}
\multiput(884.00,326.92)(1.298,-0.496){39}{\rule{1.129pt}{0.119pt}}
\multiput(884.00,327.17)(51.658,-21.000){2}{\rule{0.564pt}{0.400pt}}
\multiput(938.00,305.92)(1.410,-0.495){35}{\rule{1.216pt}{0.119pt}}
\multiput(938.00,306.17)(50.477,-19.000){2}{\rule{0.608pt}{0.400pt}}
\multiput(991.00,286.92)(1.519,-0.495){33}{\rule{1.300pt}{0.119pt}}
\multiput(991.00,287.17)(51.302,-18.000){2}{\rule{0.650pt}{0.400pt}}
\multiput(1045.00,268.92)(1.682,-0.494){29}{\rule{1.425pt}{0.119pt}}
\multiput(1045.00,269.17)(50.042,-16.000){2}{\rule{0.712pt}{0.400pt}}
\multiput(1098.00,252.92)(1.714,-0.494){29}{\rule{1.450pt}{0.119pt}}
\multiput(1098.00,253.17)(50.990,-16.000){2}{\rule{0.725pt}{0.400pt}}
\multiput(1152.00,236.92)(1.831,-0.494){27}{\rule{1.540pt}{0.119pt}}
\multiput(1152.00,237.17)(50.804,-15.000){2}{\rule{0.770pt}{0.400pt}}
\multiput(1206.00,221.92)(1.929,-0.494){25}{\rule{1.614pt}{0.119pt}}
\multiput(1206.00,222.17)(49.649,-14.000){2}{\rule{0.807pt}{0.400pt}}
\multiput(1259.00,207.92)(2.122,-0.493){23}{\rule{1.762pt}{0.119pt}}
\multiput(1259.00,208.17)(50.344,-13.000){2}{\rule{0.881pt}{0.400pt}}
\multiput(1313.00,194.94)(2.382,-0.468){5}{\rule{1.800pt}{0.113pt}}
\multiput(1313.00,195.17)(13.264,-4.000){2}{\rule{0.900pt}{0.400pt}}
\end{Large}
\end{picture}
\vskip-0.6cm
\caption{Field intensity over $\mu^4$ as a function of $k_t$ 
for $\mu=200 {\rm MeV}$ (squares),
$\mu=100 {\rm MeV}$ (pluses), and $\mu=50 {\rm MeV}$ (diamonds). Solid line 
is the LPTh prediction. The field intensity is in arbitrary units and
$k_t$ is in GeV.}
\label{na2vskl160}
\end{figure}

Some of the quantities we can compute on the lattice in strong coupling and 
in principle compare to experiment are quantities related to various 
components of the energy--momentum tensor. In an interesting recent 
paper, Testa has shown~\cite{Testa} that a class of semi--inclusive 
observables (such as, for example, the energy flux through a surface 
corresponding to the resolution of a detector) 
can be related to various components of the energy--momentum
tensor. Our eventual goal is to compute in the classical effective theory, 
non--perturbative corrections to the energy--energy correlators measured 
in jet physics~\cite{Brownetal}.

In this work, we measure a variety of observables. 
On the one hand, extensive quantities like the energy,
with well defined local gauge-invariant densities, provide useful
information on the system and are easily measurable.
On the other hand, in an experiment one usually deals with gluon
fields with definite momenta. Such fields are not easy to define
outside perturbation theory. In order to study momentum distributions
of gluons, we introduce two types of quantities.

First, we define the gauge field on every link of the lattice as the
imaginary part of the link matrix. Note that we work in the 
$A_\tau=0$ gauge and fix the Coulomb gauge in the transverse plane for
$\tau=0$. Hence there is no gauge freedom left in our definition
of the gauge field. The same applies for the scalar field. In the following,
we will study the momentum dependence of these gauge-fixed objects.

Secondly, we consider a gauge-invariant estimate of the gluon energy in 
terms of equal time energy--energy correlators.
We make use of the fact that each of the kinetic and potential terms in 
(\ref{hl}) are squared quantities such as $E^2$ and 
$B^2$. Consider, for instance, the connected correlator in the Abelian case
\be
C_m(x_t)=\langle B^2(x_t) B^2(0)\rangle_\rho-
\langle B^2(x_t)\rangle_\rho\langle B^2(0)\rangle_\rho \, .
\ee
If the distribution of $B$ were normal, $P[B]\propto\exp(h[B])$, with a
translation-invariant bilinear functional $h$, $C_m(x_t)$ would factorize:
\be
C_m(x_t)=2\langle B(x_t)B (0)\rangle_\rho^2\geq 0.
\ee
Following the analogy with this special Abelian case, we {\it define} 
the magnetic energy of the momentum mode $k_t$ in the non--Abelian theory as
\be
E_m(k_t)\equiv{1\over{2\sqrt{N}}}\sum_{\bf x_t}{\rm e}^{i{\bf{k_t\cdot x_t}}}
\sqrt{{1\over 2}C(x_t)}.
\label{emodef}\ee
The analog of (\ref{emodef}) can be defined for each of the six potential and
kinetic terms (counting the chromo-electric and the scalar potential
terms separately for each lattice direction) in (\ref{hl}). The total 
energy distribution in momentum space is then defined as the sum of these six 
contributions.

Before we end this section, we should comment on the continuum limit
in this approach. The theory contains dimensionful quantities, $\mu$
and $L$, which may be related to physical observables. One approaches
the continuum limit by keeping $\mu L$ fixed and taking $\mu$ in
lattice units to zero.  Comparing dimensionless ratios of physical
observables to their lattice counterparts would then allow one to extract
the continuum value of $\mu L$.  In this theory, the field amplitudes
squared fall off as $1/k^4$ as opposed to a thermal field theory where
the field amplitudes squared fall off as $1/k^2$. It is therefore more
likely that a wider range of physical observables will have a well
defined continuum limit than in the thermal case. Whether this is
indeed the case requires a careful study of physical (gauge invariant)
observables by taking the continuum limit of their lattice
counterparts in the manner prescribed above. We plan to continue this study 
in a future work.

Finally, we mention here for future reference a proper-time
lattice computation of an interesting gauge-invariant quantity that can 
be directly related to experiment, namely,
the Poynting vector. The Poynting vector describes the energy flux out of 
the nuclear surface. It, and other components of the energy--momentum 
tensor, can be defined on the lattice~\cite{KarschWyld}. Computing the 
Poynting vector however requires a more careful choice of boundary conditions 
than the periodic boundary conditions described in this work. This too will 
be left to a future study.

\section{Real time lattice computation of gluon production: results}
\vspace*{0.3cm}

We now turn to numerical results from our simulations. These were
performed for a variety of lattice sizes $L$= 20--160 and the color
charge density $\mu=0.015$--$0.2$ in units of the lattice spacing
$a$. To convert lattice results to physical units, we take $L^2
\approx \pi R^2$ for $A$=200 nuclei. This then also determines $\mu$
for a fixed lattice size. The relation of lattice time to continuum
time is given by the relation $\tau_C = a\cdot
\tau_L$~\cite{Krasnitz}.  In Fig.~\ref{ekvsl}, we plot the Gaussian
averaged initial kinetic energy $\langle E_k\rangle$ on the lattice as
a function of the lattice size $L$ and compare it with the lattice
perturbation theory expression (Eq.~\ref{dike} in appendix B) for
different values $\mu_L = 0.0177, 0.035$ of the color charge
density. For small values of $L$, there is very good agreement between
the two but for the largest value $L=160$, they begin to
deviate. Since the strong coupling parameter on the lattice is
$\propto g^2\mu_L L$, for $g^2 \mu_L L \gg 1$ we can expect to see
deviations from lattice perturbation theory.

\begin{figure}[ht]
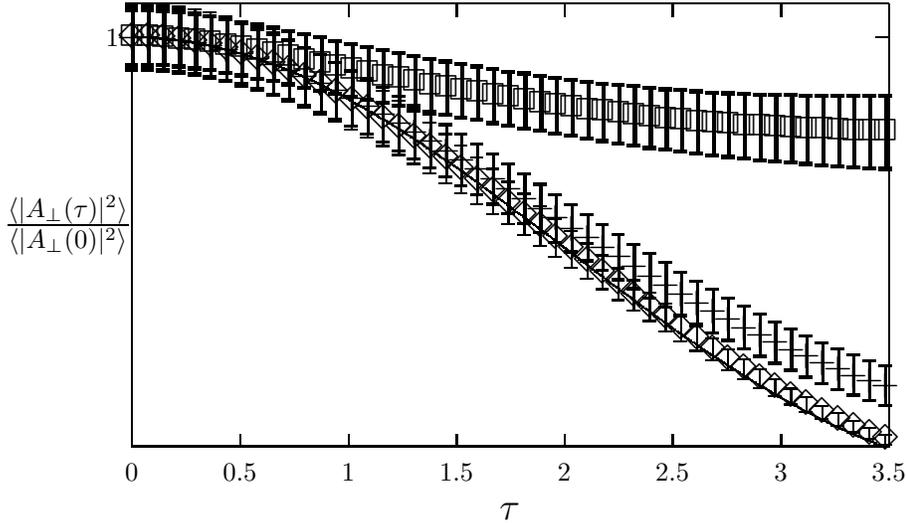

\setlength{\unitlength}{0.240900pt}
\ifx\plotpoint\undefined\newsavebox{\plotpoint}\fi
\sbox{\plotpoint}{\rule[-0.200pt]{0.400pt}{0.400pt}}%

\vskip-0.6cm
\caption{Normalized field intensity of a soft ($k_t=108{\rm MeV}$) mode 
vs proper time $\tau$ (in units of fm) for $\mu=200{\rm MeV}$ (squares),
$\mu=100{\rm MeV}$ (pluses), and $\mu=50{\rm MeV}$ (diamonds). Solid line, 
nearly coinciding with the $\mu=50{\rm MeV}$ curve, is the LPTh prediction.}
\label{a2kvsphtminkl160}
\end{figure}

In Fig.~\ref{a2kvspht025pil160mu025}, we plot the ratio of the field 
intensity of a particular
mode of the transverse gauge field as a function of proper time $\tau$
normalized to its value at $\tau=0$. The diamonds are results from a
lattice simulation with $L=160$ and $\mu_L =0.0177$ and the mode
considered is $(k_x,k_y)=(\pi/4,0)$. Note that $k_{x,y}=2\pi
n_{x,y}/L$ and for this case, $n_x =20, n_y=0$. The solid line in the
figure is the square of the Bessel function $J_0 (\omega \tau)$ where
$\omega = \sqrt{2 (2 -\cos(k_x)-\cos(k_y))}$. 
Since the $\tau$--direction in our simulation is continuous, the time
dependence of the high transverse momentum perturbative modes should
agree with the continuum perturbative result 
which predicts a time dependence proportional to $J_0^2 (\omega\tau)$ 
for the field
intensity of the transverse gauge fields. The continuum dispersion relation
$\omega =|k_\perp|$ is however modified into the lattice dispersion relation
shown above. 
We see from the figure that the anticipated agreement
between the lattice results and perturbation theory is quite good.

\begin{figure}[ht]
\setlength{\unitlength}{0.240900pt}
\ifx\plotpoint\undefined\newsavebox{\plotpoint}\fi
\sbox{\plotpoint}{\rule[-0.200pt]{0.400pt}{0.400pt}}%
\begin{picture}(1350,900)(0,0)
\begin{Large}
\font\gnuplot=cmr10 at 10pt
\gnuplot
\sbox{\plotpoint}{\rule[-0.200pt]{0.400pt}{0.400pt}}%
\put(141.0,163.0){\rule[-0.200pt]{4.818pt}{0.400pt}}
\put(121,163){\makebox(0,0)[r]{0}}
\put(1310.0,163.0){\rule[-0.200pt]{4.818pt}{0.400pt}}
\put(141.0,472.0){\rule[-0.200pt]{4.818pt}{0.400pt}}
\put(121,472){\makebox(0,0)[r]{2}}
\put(1310.0,472.0){\rule[-0.200pt]{4.818pt}{0.400pt}}
\put(141.0,782.0){\rule[-0.200pt]{4.818pt}{0.400pt}}
\put(121,782){\makebox(0,0)[r]{4}}
\put(1310.0,782.0){\rule[-0.200pt]{4.818pt}{0.400pt}}
\put(141.0,163.0){\rule[-0.200pt]{0.400pt}{4.818pt}}
\put(141,122){\makebox(0,0){0}}
\put(141.0,839.0){\rule[-0.200pt]{0.400pt}{4.818pt}}
\put(379.0,163.0){\rule[-0.200pt]{0.400pt}{4.818pt}}
\put(379,122){\makebox(0,0){1}}
\put(379.0,839.0){\rule[-0.200pt]{0.400pt}{4.818pt}}
\put(617.0,163.0){\rule[-0.200pt]{0.400pt}{4.818pt}}
\put(617,122){\makebox(0,0){2}}
\put(617.0,839.0){\rule[-0.200pt]{0.400pt}{4.818pt}}
\put(854.0,163.0){\rule[-0.200pt]{0.400pt}{4.818pt}}
\put(854,122){\makebox(0,0){3}}
\put(854.0,839.0){\rule[-0.200pt]{0.400pt}{4.818pt}}
\put(1092.0,163.0){\rule[-0.200pt]{0.400pt}{4.818pt}}
\put(1092,122){\makebox(0,0){4}}
\put(1092.0,839.0){\rule[-0.200pt]{0.400pt}{4.818pt}}
\put(1330.0,163.0){\rule[-0.200pt]{0.400pt}{4.818pt}}
\put(1330,122){\makebox(0,0){5}}
\put(1330.0,839.0){\rule[-0.200pt]{0.400pt}{4.818pt}}
\put(141.0,163.0){\rule[-0.200pt]{286.430pt}{0.400pt}}
\put(1330.0,163.0){\rule[-0.200pt]{0.400pt}{167.666pt}}
\put(141.0,859.0){\rule[-0.200pt]{286.430pt}{0.400pt}}
\put(41,511){\makebox(0,0){$\tau E$}}
\put(735,61){\makebox(0,0){$\tau$}}
\put(141.0,163.0){\rule[-0.200pt]{0.400pt}{167.666pt}}
\put(141,163){\raisebox{-.8pt}{\makebox(0,0){$\Diamond$}}}
\put(158,351){\raisebox{-.8pt}{\makebox(0,0){$\Diamond$}}}
\put(175,441){\raisebox{-.8pt}{\makebox(0,0){$\Diamond$}}}
\put(193,501){\raisebox{-.8pt}{\makebox(0,0){$\Diamond$}}}
\put(210,545){\raisebox{-.8pt}{\makebox(0,0){$\Diamond$}}}
\put(227,578){\raisebox{-.8pt}{\makebox(0,0){$\Diamond$}}}
\put(244,603){\raisebox{-.8pt}{\makebox(0,0){$\Diamond$}}}
\put(262,623){\raisebox{-.8pt}{\makebox(0,0){$\Diamond$}}}
\put(279,639){\raisebox{-.8pt}{\makebox(0,0){$\Diamond$}}}
\put(296,652){\raisebox{-.8pt}{\makebox(0,0){$\Diamond$}}}
\put(313,662){\raisebox{-.8pt}{\makebox(0,0){$\Diamond$}}}
\put(331,671){\raisebox{-.8pt}{\makebox(0,0){$\Diamond$}}}
\put(348,678){\raisebox{-.8pt}{\makebox(0,0){$\Diamond$}}}
\put(365,684){\raisebox{-.8pt}{\makebox(0,0){$\Diamond$}}}
\put(382,688){\raisebox{-.8pt}{\makebox(0,0){$\Diamond$}}}
\put(400,692){\raisebox{-.8pt}{\makebox(0,0){$\Diamond$}}}
\put(417,696){\raisebox{-.8pt}{\makebox(0,0){$\Diamond$}}}
\put(434,698){\raisebox{-.8pt}{\makebox(0,0){$\Diamond$}}}
\put(451,701){\raisebox{-.8pt}{\makebox(0,0){$\Diamond$}}}
\put(469,703){\raisebox{-.8pt}{\makebox(0,0){$\Diamond$}}}
\put(486,704){\raisebox{-.8pt}{\makebox(0,0){$\Diamond$}}}
\put(503,706){\raisebox{-.8pt}{\makebox(0,0){$\Diamond$}}}
\put(520,707){\raisebox{-.8pt}{\makebox(0,0){$\Diamond$}}}
\put(538,708){\raisebox{-.8pt}{\makebox(0,0){$\Diamond$}}}
\put(555,708){\raisebox{-.8pt}{\makebox(0,0){$\Diamond$}}}
\put(572,709){\raisebox{-.8pt}{\makebox(0,0){$\Diamond$}}}
\put(589,710){\raisebox{-.8pt}{\makebox(0,0){$\Diamond$}}}
\put(606,710){\raisebox{-.8pt}{\makebox(0,0){$\Diamond$}}}
\put(624,711){\raisebox{-.8pt}{\makebox(0,0){$\Diamond$}}}
\put(641,711){\raisebox{-.8pt}{\makebox(0,0){$\Diamond$}}}
\put(658,711){\raisebox{-.8pt}{\makebox(0,0){$\Diamond$}}}
\put(675,712){\raisebox{-.8pt}{\makebox(0,0){$\Diamond$}}}
\put(693,712){\raisebox{-.8pt}{\makebox(0,0){$\Diamond$}}}
\put(710,712){\raisebox{-.8pt}{\makebox(0,0){$\Diamond$}}}
\put(727,713){\raisebox{-.8pt}{\makebox(0,0){$\Diamond$}}}
\put(744,713){\raisebox{-.8pt}{\makebox(0,0){$\Diamond$}}}
\put(762,713){\raisebox{-.8pt}{\makebox(0,0){$\Diamond$}}}
\put(779,714){\raisebox{-.8pt}{\makebox(0,0){$\Diamond$}}}
\put(796,714){\raisebox{-.8pt}{\makebox(0,0){$\Diamond$}}}
\put(813,714){\raisebox{-.8pt}{\makebox(0,0){$\Diamond$}}}
\put(831,715){\raisebox{-.8pt}{\makebox(0,0){$\Diamond$}}}
\put(848,715){\raisebox{-.8pt}{\makebox(0,0){$\Diamond$}}}
\put(865,715){\raisebox{-.8pt}{\makebox(0,0){$\Diamond$}}}
\put(882,716){\raisebox{-.8pt}{\makebox(0,0){$\Diamond$}}}
\put(900,716){\raisebox{-.8pt}{\makebox(0,0){$\Diamond$}}}
\put(917,717){\raisebox{-.8pt}{\makebox(0,0){$\Diamond$}}}
\put(934,717){\raisebox{-.8pt}{\makebox(0,0){$\Diamond$}}}
\put(951,717){\raisebox{-.8pt}{\makebox(0,0){$\Diamond$}}}
\put(969,718){\raisebox{-.8pt}{\makebox(0,0){$\Diamond$}}}
\put(986,718){\raisebox{-.8pt}{\makebox(0,0){$\Diamond$}}}
\put(1003,719){\raisebox{-.8pt}{\makebox(0,0){$\Diamond$}}}
\put(1020,719){\raisebox{-.8pt}{\makebox(0,0){$\Diamond$}}}
\put(1038,720){\raisebox{-.8pt}{\makebox(0,0){$\Diamond$}}}
\put(1055,720){\raisebox{-.8pt}{\makebox(0,0){$\Diamond$}}}
\put(1072,721){\raisebox{-.8pt}{\makebox(0,0){$\Diamond$}}}
\put(1089,721){\raisebox{-.8pt}{\makebox(0,0){$\Diamond$}}}
\put(1106,722){\raisebox{-.8pt}{\makebox(0,0){$\Diamond$}}}
\put(1124,722){\raisebox{-.8pt}{\makebox(0,0){$\Diamond$}}}
\put(1141,723){\raisebox{-.8pt}{\makebox(0,0){$\Diamond$}}}
\put(1158,723){\raisebox{-.8pt}{\makebox(0,0){$\Diamond$}}}
\put(1175,724){\raisebox{-.8pt}{\makebox(0,0){$\Diamond$}}}
\put(1193,725){\raisebox{-.8pt}{\makebox(0,0){$\Diamond$}}}
\put(1210,725){\raisebox{-.8pt}{\makebox(0,0){$\Diamond$}}}
\put(1227,726){\raisebox{-.8pt}{\makebox(0,0){$\Diamond$}}}
\put(1244,726){\raisebox{-.8pt}{\makebox(0,0){$\Diamond$}}}
\put(1262,727){\raisebox{-.8pt}{\makebox(0,0){$\Diamond$}}}
\put(1279,727){\raisebox{-.8pt}{\makebox(0,0){$\Diamond$}}}
\put(1296,728){\raisebox{-.8pt}{\makebox(0,0){$\Diamond$}}}
\put(1313,729){\raisebox{-.8pt}{\makebox(0,0){$\Diamond$}}}
\sbox{\plotpoint}{\rule[-0.400pt]{0.800pt}{0.800pt}}%
\put(141,163){\makebox(0,0){$+$}}
\put(158,336){\makebox(0,0){$+$}}
\put(175,419){\makebox(0,0){$+$}}
\put(193,473){\makebox(0,0){$+$}}
\put(210,512){\makebox(0,0){$+$}}
\put(227,539){\makebox(0,0){$+$}}
\put(244,560){\makebox(0,0){$+$}}
\put(262,575){\makebox(0,0){$+$}}
\put(279,586){\makebox(0,0){$+$}}
\put(296,595){\makebox(0,0){$+$}}
\put(313,601){\makebox(0,0){$+$}}
\put(331,606){\makebox(0,0){$+$}}
\put(348,610){\makebox(0,0){$+$}}
\put(365,613){\makebox(0,0){$+$}}
\put(382,615){\makebox(0,0){$+$}}
\put(400,616){\makebox(0,0){$+$}}
\put(417,618){\makebox(0,0){$+$}}
\put(434,619){\makebox(0,0){$+$}}
\put(451,619){\makebox(0,0){$+$}}
\put(469,620){\makebox(0,0){$+$}}
\put(486,620){\makebox(0,0){$+$}}
\put(503,621){\makebox(0,0){$+$}}
\put(520,621){\makebox(0,0){$+$}}
\put(538,621){\makebox(0,0){$+$}}
\put(555,621){\makebox(0,0){$+$}}
\put(572,622){\makebox(0,0){$+$}}
\put(589,622){\makebox(0,0){$+$}}
\put(606,622){\makebox(0,0){$+$}}
\put(624,622){\makebox(0,0){$+$}}
\put(641,622){\makebox(0,0){$+$}}
\put(658,622){\makebox(0,0){$+$}}
\put(675,622){\makebox(0,0){$+$}}
\put(693,622){\makebox(0,0){$+$}}
\put(710,622){\makebox(0,0){$+$}}
\put(727,622){\makebox(0,0){$+$}}
\put(744,623){\makebox(0,0){$+$}}
\put(762,623){\makebox(0,0){$+$}}
\put(779,623){\makebox(0,0){$+$}}
\put(796,623){\makebox(0,0){$+$}}
\put(813,623){\makebox(0,0){$+$}}
\put(831,623){\makebox(0,0){$+$}}
\put(848,624){\makebox(0,0){$+$}}
\put(865,624){\makebox(0,0){$+$}}
\put(882,624){\makebox(0,0){$+$}}
\put(900,624){\makebox(0,0){$+$}}
\put(917,624){\makebox(0,0){$+$}}
\put(934,624){\makebox(0,0){$+$}}
\put(951,624){\makebox(0,0){$+$}}
\put(969,625){\makebox(0,0){$+$}}
\put(986,625){\makebox(0,0){$+$}}
\put(1003,625){\makebox(0,0){$+$}}
\put(1020,625){\makebox(0,0){$+$}}
\put(1038,625){\makebox(0,0){$+$}}
\put(1055,625){\makebox(0,0){$+$}}
\put(1072,625){\makebox(0,0){$+$}}
\put(1089,625){\makebox(0,0){$+$}}
\put(1106,625){\makebox(0,0){$+$}}
\put(1124,625){\makebox(0,0){$+$}}
\put(1141,625){\makebox(0,0){$+$}}
\put(1158,625){\makebox(0,0){$+$}}
\put(1175,625){\makebox(0,0){$+$}}
\put(1193,625){\makebox(0,0){$+$}}
\put(1210,625){\makebox(0,0){$+$}}
\put(1227,625){\makebox(0,0){$+$}}
\put(1244,625){\makebox(0,0){$+$}}
\put(1262,625){\makebox(0,0){$+$}}
\put(1279,625){\makebox(0,0){$+$}}
\put(1296,625){\makebox(0,0){$+$}}
\put(1313,625){\makebox(0,0){$+$}}
\sbox{\plotpoint}{\rule[-0.500pt]{1.000pt}{1.000pt}}%
\sbox{\plotpoint}{\rule[-0.600pt]{1.200pt}{1.200pt}}%
\put(141,163){\raisebox{-.8pt}{\makebox(0,0){$\Box$}}}
\put(158,276){\raisebox{-.8pt}{\makebox(0,0){$\Box$}}}
\put(175,326){\raisebox{-.8pt}{\makebox(0,0){$\Box$}}}
\put(193,356){\raisebox{-.8pt}{\makebox(0,0){$\Box$}}}
\put(210,374){\raisebox{-.8pt}{\makebox(0,0){$\Box$}}}
\put(227,385){\raisebox{-.8pt}{\makebox(0,0){$\Box$}}}
\put(244,391){\raisebox{-.8pt}{\makebox(0,0){$\Box$}}}
\put(262,395){\raisebox{-.8pt}{\makebox(0,0){$\Box$}}}
\put(279,398){\raisebox{-.8pt}{\makebox(0,0){$\Box$}}}
\put(296,399){\raisebox{-.8pt}{\makebox(0,0){$\Box$}}}
\put(313,400){\raisebox{-.8pt}{\makebox(0,0){$\Box$}}}
\put(331,400){\raisebox{-.8pt}{\makebox(0,0){$\Box$}}}
\put(348,400){\raisebox{-.8pt}{\makebox(0,0){$\Box$}}}
\put(365,400){\raisebox{-.8pt}{\makebox(0,0){$\Box$}}}
\put(382,400){\raisebox{-.8pt}{\makebox(0,0){$\Box$}}}
\put(400,400){\raisebox{-.8pt}{\makebox(0,0){$\Box$}}}
\put(417,399){\raisebox{-.8pt}{\makebox(0,0){$\Box$}}}
\put(434,399){\raisebox{-.8pt}{\makebox(0,0){$\Box$}}}
\put(451,399){\raisebox{-.8pt}{\makebox(0,0){$\Box$}}}
\put(469,399){\raisebox{-.8pt}{\makebox(0,0){$\Box$}}}
\put(486,399){\raisebox{-.8pt}{\makebox(0,0){$\Box$}}}
\put(503,398){\raisebox{-.8pt}{\makebox(0,0){$\Box$}}}
\put(520,398){\raisebox{-.8pt}{\makebox(0,0){$\Box$}}}
\put(538,398){\raisebox{-.8pt}{\makebox(0,0){$\Box$}}}
\put(555,398){\raisebox{-.8pt}{\makebox(0,0){$\Box$}}}
\put(572,398){\raisebox{-.8pt}{\makebox(0,0){$\Box$}}}
\put(589,397){\raisebox{-.8pt}{\makebox(0,0){$\Box$}}}
\put(606,397){\raisebox{-.8pt}{\makebox(0,0){$\Box$}}}
\put(624,397){\raisebox{-.8pt}{\makebox(0,0){$\Box$}}}
\put(641,397){\raisebox{-.8pt}{\makebox(0,0){$\Box$}}}
\put(658,397){\raisebox{-.8pt}{\makebox(0,0){$\Box$}}}
\put(675,397){\raisebox{-.8pt}{\makebox(0,0){$\Box$}}}
\put(693,396){\raisebox{-.8pt}{\makebox(0,0){$\Box$}}}
\put(710,396){\raisebox{-.8pt}{\makebox(0,0){$\Box$}}}
\put(727,396){\raisebox{-.8pt}{\makebox(0,0){$\Box$}}}
\put(744,396){\raisebox{-.8pt}{\makebox(0,0){$\Box$}}}
\put(762,396){\raisebox{-.8pt}{\makebox(0,0){$\Box$}}}
\put(779,395){\raisebox{-.8pt}{\makebox(0,0){$\Box$}}}
\put(796,395){\raisebox{-.8pt}{\makebox(0,0){$\Box$}}}
\put(813,395){\raisebox{-.8pt}{\makebox(0,0){$\Box$}}}
\put(831,395){\raisebox{-.8pt}{\makebox(0,0){$\Box$}}}
\put(848,395){\raisebox{-.8pt}{\makebox(0,0){$\Box$}}}
\put(865,394){\raisebox{-.8pt}{\makebox(0,0){$\Box$}}}
\put(882,394){\raisebox{-.8pt}{\makebox(0,0){$\Box$}}}
\put(900,394){\raisebox{-.8pt}{\makebox(0,0){$\Box$}}}
\put(917,394){\raisebox{-.8pt}{\makebox(0,0){$\Box$}}}
\put(934,394){\raisebox{-.8pt}{\makebox(0,0){$\Box$}}}
\put(951,393){\raisebox{-.8pt}{\makebox(0,0){$\Box$}}}
\put(969,393){\raisebox{-.8pt}{\makebox(0,0){$\Box$}}}
\put(986,393){\raisebox{-.8pt}{\makebox(0,0){$\Box$}}}
\put(1003,393){\raisebox{-.8pt}{\makebox(0,0){$\Box$}}}
\put(1020,393){\raisebox{-.8pt}{\makebox(0,0){$\Box$}}}
\put(1038,393){\raisebox{-.8pt}{\makebox(0,0){$\Box$}}}
\put(1055,392){\raisebox{-.8pt}{\makebox(0,0){$\Box$}}}
\put(1072,392){\raisebox{-.8pt}{\makebox(0,0){$\Box$}}}
\put(1089,392){\raisebox{-.8pt}{\makebox(0,0){$\Box$}}}
\put(1106,392){\raisebox{-.8pt}{\makebox(0,0){$\Box$}}}
\put(1124,392){\raisebox{-.8pt}{\makebox(0,0){$\Box$}}}
\put(1141,392){\raisebox{-.8pt}{\makebox(0,0){$\Box$}}}
\put(1158,392){\raisebox{-.8pt}{\makebox(0,0){$\Box$}}}
\put(1175,392){\raisebox{-.8pt}{\makebox(0,0){$\Box$}}}
\put(1193,391){\raisebox{-.8pt}{\makebox(0,0){$\Box$}}}
\put(1210,391){\raisebox{-.8pt}{\makebox(0,0){$\Box$}}}
\put(1227,391){\raisebox{-.8pt}{\makebox(0,0){$\Box$}}}
\put(1244,391){\raisebox{-.8pt}{\makebox(0,0){$\Box$}}}
\put(1262,391){\raisebox{-.8pt}{\makebox(0,0){$\Box$}}}
\put(1279,391){\raisebox{-.8pt}{\makebox(0,0){$\Box$}}}
\put(1296,390){\raisebox{-.8pt}{\makebox(0,0){$\Box$}}}
\put(1313,390){\raisebox{-.8pt}{\makebox(0,0){$\Box$}}}
\end{Large}
\end{picture}
\vskip-0.6cm
\caption{Time history of the energy density in units of $\mu^4$
for $\mu=200{\rm MeV}$ (squares),
$\mu=100{\rm MeV}$ (pluses), and $\mu=50{\rm MeV}$ (diamonds).
Error bars are smaller than the plotting symbols. Proper time $\tau$ is
in fm.}
\label{ehistl160}
\end{figure}
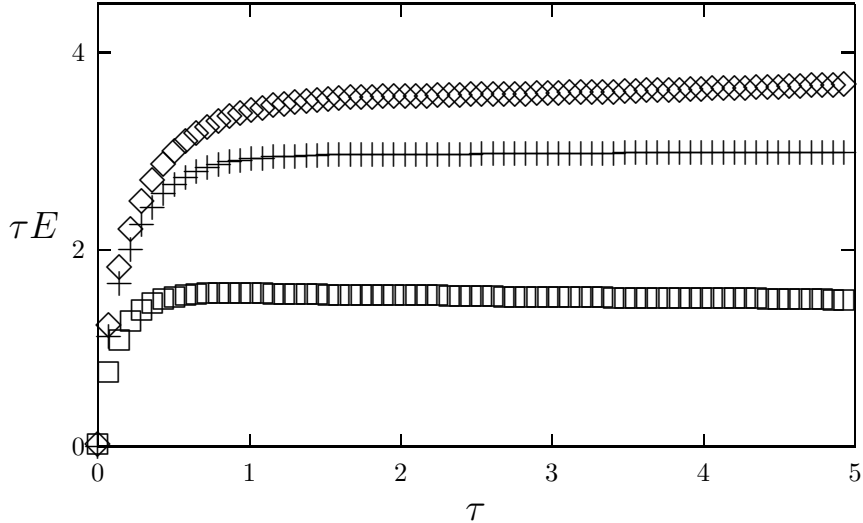

In Fig.~\ref{na2vskl160}, the field intensity of the transverse gauge field 
normalized by $\mu^4$ at
$\tau =0$ is plotted as a function of the transverse momentum in 
physical units.  The lattice results for the different values of $\mu$ 
described in the caption
are compared to lattice perturbation theory result--Eq.~\ref{dcrfin} 
in appendix B--for the field intensity.
The LPTh result (which would be the mini--jet distribution in the 
continuum) agrees very well with the lattice result for small $\mu$ upto
very small values of $k_t$. However, strong coupling effects grow with 
increasing $\mu$ (the lattice size $L$ is fixed here) and we see deviations
from the perturbative predictions at larger values of $k_t$. This trend is
enhanced further at larger values of $\mu$ than those shown here. The 
non--perturbative effects due to the non--linearities in the Yang--Mills 
equations seem to temper the $1/k_t^4$ behaviour predicted by perturbation 
theory. Whether this reflects the presence of a 
mass in the theory needs further investigation.

In Fig.~\ref{a2kvsphtminkl160}, we plot the same quantity as in Fig.~2, 
but now for three different 
values of $\mu$ and for the first non--zero momentum mode $(k_x,k_y) = (1,0)$.
The lattice size $L=160$ is the same as previously. For the smallest value of
$\mu = 0.0177$, there is again an agreement with the Bessel behaviour predicted
by perturbation theory. However at the larger values of 
$\mu_L = 0.035, 0.07$, one
sees significant deviations away from the Bessel behaviour. Indeed, the 
modes seem to saturate at larger values of $\tau$. At face value
this is unexpected because one expects that as the systems cools at late 
times, even the small $k_\perp$ modes should eventually die out. It 
may be that we have to wait for times much longer than those studied to 
see this. That would indeed be very interesting because these times would be
much greater than the natural time scale $\tau\sim 1/g^2 \mu$ at which we
expect non--linearities to dissipate (see the discussion of 
Fig.~\ref{ehistl160} below).

\begin{figure}[ht]
\setlength{\unitlength}{0.240900pt}
\ifx\plotpoint\undefined\newsavebox{\plotpoint}\fi
\sbox{\plotpoint}{\rule[-0.200pt]{0.400pt}{0.400pt}}%
\begin{picture}(1350,900)(0,0)
\begin{Large}
\font\gnuplot=cmr10 at 10pt
\gnuplot
\sbox{\plotpoint}{\rule[-0.200pt]{0.400pt}{0.400pt}}%
\put(181.0,163.0){\rule[-0.200pt]{4.818pt}{0.400pt}}
\put(161,163){\makebox(0,0)[r]{0}}
\put(1310.0,163.0){\rule[-0.200pt]{4.818pt}{0.400pt}}
\put(181.0,511.0){\rule[-0.200pt]{4.818pt}{0.400pt}}
\put(161,511){\makebox(0,0)[r]{0.5}}
\put(1310.0,511.0){\rule[-0.200pt]{4.818pt}{0.400pt}}
\put(181.0,859.0){\rule[-0.200pt]{4.818pt}{0.400pt}}
\put(161,859){\makebox(0,0)[r]{1}}
\put(1310.0,859.0){\rule[-0.200pt]{4.818pt}{0.400pt}}
\put(181.0,163.0){\rule[-0.200pt]{0.400pt}{4.818pt}}
\put(181,122){\makebox(0,0){0}}
\put(181.0,839.0){\rule[-0.200pt]{0.400pt}{4.818pt}}
\put(564.0,163.0){\rule[-0.200pt]{0.400pt}{4.818pt}}
\put(564,122){\makebox(0,0){2}}
\put(564.0,839.0){\rule[-0.200pt]{0.400pt}{4.818pt}}
\put(947.0,163.0){\rule[-0.200pt]{0.400pt}{4.818pt}}
\put(947,122){\makebox(0,0){4}}
\put(947.0,839.0){\rule[-0.200pt]{0.400pt}{4.818pt}}
\put(1330.0,163.0){\rule[-0.200pt]{0.400pt}{4.818pt}}
\put(1330,122){\makebox(0,0){6}}
\put(1330.0,839.0){\rule[-0.200pt]{0.400pt}{4.818pt}}
\put(181.0,163.0){\rule[-0.200pt]{276.794pt}{0.400pt}}
\put(1330.0,163.0){\rule[-0.200pt]{0.400pt}{167.666pt}}
\put(181.0,859.0){\rule[-0.200pt]{276.794pt}{0.400pt}}
\put(41,511){\makebox(0,0){\Large${{\langle E_k\rangle}\over{\langle E\rangle}}$}}
\put(755,61){\makebox(0,0){\Large$\tau$}}
\put(181.0,163.0){\rule[-0.200pt]{0.400pt}{167.666pt}}
\put(181,700){\usebox{\plotpoint}}
\multiput(181.00,698.93)(3.531,-0.488){13}{\rule{2.800pt}{0.117pt}}
\multiput(181.00,699.17)(48.188,-8.000){2}{\rule{1.400pt}{0.400pt}}
\multiput(235.00,692.58)(0.918,0.497){55}{\rule{0.831pt}{0.120pt}}
\multiput(235.00,691.17)(51.275,29.000){2}{\rule{0.416pt}{0.400pt}}
\multiput(288.00,721.59)(3.116,0.489){15}{\rule{2.500pt}{0.118pt}}
\multiput(288.00,720.17)(48.811,9.000){2}{\rule{1.250pt}{0.400pt}}
\multiput(342.58,727.61)(0.498,-0.594){103}{\rule{0.120pt}{0.575pt}}
\multiput(341.17,728.81)(53.000,-61.806){2}{\rule{0.400pt}{0.288pt}}
\multiput(395.58,663.36)(0.498,-0.975){105}{\rule{0.120pt}{0.878pt}}
\multiput(394.17,665.18)(54.000,-103.178){2}{\rule{0.400pt}{0.439pt}}
\multiput(449.58,559.50)(0.499,-0.627){265}{\rule{0.120pt}{0.601pt}}
\multiput(448.17,560.75)(134.000,-166.752){2}{\rule{0.400pt}{0.301pt}}
\multiput(583.00,394.58)(5.294,0.493){23}{\rule{4.223pt}{0.119pt}}
\multiput(583.00,393.17)(125.235,13.000){2}{\rule{2.112pt}{0.400pt}}
\multiput(717.00,405.92)(2.362,-0.499){111}{\rule{1.981pt}{0.120pt}}
\multiput(717.00,406.17)(263.889,-57.000){2}{\rule{0.990pt}{0.400pt}}
\multiput(985.00,348.93)(24.177,-0.482){9}{\rule{17.967pt}{0.116pt}}
\multiput(985.00,349.17)(230.709,-6.000){2}{\rule{8.983pt}{0.400pt}}
\put(181,700){\raisebox{-.8pt}{\makebox(0,0){$\Diamond$}}}
\put(235,692){\raisebox{-.8pt}{\makebox(0,0){$\Diamond$}}}
\put(288,721){\raisebox{-.8pt}{\makebox(0,0){$\Diamond$}}}
\put(342,730){\raisebox{-.8pt}{\makebox(0,0){$\Diamond$}}}
\put(395,667){\raisebox{-.8pt}{\makebox(0,0){$\Diamond$}}}
\put(449,562){\raisebox{-.8pt}{\makebox(0,0){$\Diamond$}}}
\put(583,394){\raisebox{-.8pt}{\makebox(0,0){$\Diamond$}}}
\put(717,407){\raisebox{-.8pt}{\makebox(0,0){$\Diamond$}}}
\put(985,350){\raisebox{-.8pt}{\makebox(0,0){$\Diamond$}}}
\put(1253,344){\raisebox{-.8pt}{\makebox(0,0){$\Diamond$}}}
\sbox{\plotpoint}{\rule[-0.400pt]{0.800pt}{0.800pt}}%
\put(181,657){\usebox{\plotpoint}}
\put(181,656.34){\rule{13.009pt}{0.800pt}}
\multiput(181.00,655.34)(27.000,2.000){2}{\rule{6.504pt}{0.800pt}}
\multiput(235.00,660.41)(1.980,0.509){21}{\rule{3.229pt}{0.123pt}}
\multiput(235.00,657.34)(46.299,14.000){2}{\rule{1.614pt}{0.800pt}}
\multiput(288.00,671.06)(8.652,-0.560){3}{\rule{8.840pt}{0.135pt}}
\multiput(288.00,671.34)(35.652,-5.000){2}{\rule{4.420pt}{0.800pt}}
\multiput(343.41,663.54)(0.502,-0.546){99}{\rule{0.121pt}{1.075pt}}
\multiput(340.34,665.77)(53.000,-55.768){2}{\rule{0.800pt}{0.538pt}}
\multiput(396.41,604.00)(0.502,-0.780){101}{\rule{0.121pt}{1.444pt}}
\multiput(393.34,607.00)(54.000,-81.002){2}{\rule{0.800pt}{0.722pt}}
\multiput(450.41,521.65)(0.501,-0.529){261}{\rule{0.121pt}{1.048pt}}
\multiput(447.34,523.83)(134.000,-139.825){2}{\rule{0.800pt}{0.524pt}}
\multiput(583.00,385.40)(6.623,0.512){15}{\rule{9.945pt}{0.123pt}}
\multiput(583.00,382.34)(113.358,11.000){2}{\rule{4.973pt}{0.800pt}}
\multiput(717.00,393.09)(2.657,-0.502){95}{\rule{4.404pt}{0.121pt}}
\multiput(717.00,393.34)(258.859,-51.000){2}{\rule{2.202pt}{0.800pt}}
\multiput(985.00,342.06)(44.585,-0.560){3}{\rule{43.080pt}{0.135pt}}
\multiput(985.00,342.34)(178.585,-5.000){2}{\rule{21.540pt}{0.800pt}}
\put(181,657){\raisebox{-.8pt}{\makebox(0,0){$\Box$}}}
\put(235,659){\raisebox{-.8pt}{\makebox(0,0){$\Box$}}}
\put(288,673){\raisebox{-.8pt}{\makebox(0,0){$\Box$}}}
\put(342,668){\raisebox{-.8pt}{\makebox(0,0){$\Box$}}}
\put(395,610){\raisebox{-.8pt}{\makebox(0,0){$\Box$}}}
\put(449,526){\raisebox{-.8pt}{\makebox(0,0){$\Box$}}}
\put(583,384){\raisebox{-.8pt}{\makebox(0,0){$\Box$}}}
\put(717,395){\raisebox{-.8pt}{\makebox(0,0){$\Box$}}}
\put(985,344){\raisebox{-.8pt}{\makebox(0,0){$\Box$}}}
\put(1253,339){\raisebox{-.8pt}{\makebox(0,0){$\Box$}}}
\sbox{\plotpoint}{\rule[-0.500pt]{1.000pt}{1.000pt}}%
\sbox{\plotpoint}{\rule[-0.600pt]{1.200pt}{1.200pt}}%
\put(181,276){\usebox{\plotpoint}}
\multiput(181.00,278.24)(0.592,0.500){80}{\rule{1.740pt}{0.121pt}}
\multiput(181.00,273.51)(50.389,45.000){2}{\rule{0.870pt}{1.200pt}}
\multiput(237.24,313.55)(0.500,-0.616){96}{\rule{0.120pt}{1.794pt}}
\multiput(232.51,317.28)(53.000,-62.276){2}{\rule{1.200pt}{0.897pt}}
\multiput(290.24,248.13)(0.500,-0.558){98}{\rule{0.120pt}{1.656pt}}
\multiput(285.51,251.56)(54.000,-57.564){2}{\rule{1.200pt}{0.828pt}}
\put(342,192.01){\rule{12.768pt}{1.200pt}}
\multiput(342.00,191.51)(26.500,1.000){2}{\rule{6.384pt}{1.200pt}}
\multiput(395.00,197.24)(0.667,0.500){70}{\rule{1.920pt}{0.121pt}}
\multiput(395.00,192.51)(50.015,40.000){2}{\rule{0.960pt}{1.200pt}}
\multiput(449.00,237.24)(2.046,0.500){56}{\rule{5.173pt}{0.121pt}}
\multiput(449.00,232.51)(123.264,33.000){2}{\rule{2.586pt}{1.200pt}}
\put(717,267.01){\rule{64.561pt}{1.200pt}}
\multiput(717.00,265.51)(134.000,3.000){2}{\rule{32.281pt}{1.200pt}}
\put(985,267.01){\rule{64.561pt}{1.200pt}}
\multiput(985.00,268.51)(134.000,-3.000){2}{\rule{32.281pt}{1.200pt}}
\put(181,276){\makebox(0,0){$\triangle$}}
\put(235,321){\makebox(0,0){$\triangle$}}
\put(288,255){\makebox(0,0){$\triangle$}}
\put(342,194){\makebox(0,0){$\triangle$}}
\put(395,195){\makebox(0,0){$\triangle$}}
\put(449,235){\makebox(0,0){$\triangle$}}
\put(583,268){\makebox(0,0){$\triangle$}}
\put(717,268){\makebox(0,0){$\triangle$}}
\put(985,271){\makebox(0,0){$\triangle$}}
\put(1253,268){\makebox(0,0){$\triangle$}}
\put(583.0,268.0){\rule[-0.600pt]{32.281pt}{1.200pt}}
\end{Large}
\end{picture}
\caption{Energy per mode k normalized to the total energy as a function 
of proper time in fermis. From top to bottom, the curves correspond to
modes $k=0,108$ and $432$ MeV respectively. As in the previous figure,
$\mu=0.41$ GeV.}
\label{ecorrvstau}
\end{figure}
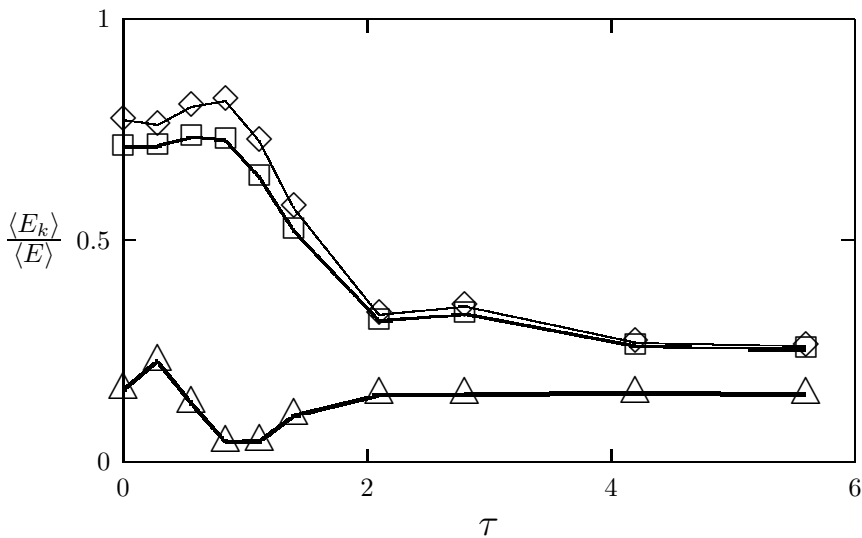

A straightforward explanation for this behaviour is that it is some kind of
gauge artifact. This is because even though the fields satisfy the Coulomb 
gauge condition at $\tau=0$, they no longer do so at later times. Thus one
may need to fix Coulomb gauge at each ``measured'' time to properly interpret
what we have called as field intensities as such. Large modes may be
unaffected by the gauge fixing but small modes will be affected. The simplest
test of this explanation is to study the correlators of gauge invariant 
quantities at late times. If they display the same ``saturation'' then the
effect cannot be dismissed as a gauge artifact. We will return to this
point a little later.

Before we do that, we would like to discuss the time dependence of the 
energy density, shown in Fig.~\ref{ehistl160} for different
values of $\mu$. At late times, from general considerations 
we expect that $E\propto 1/\tau$ and that is what we see. Indeed,
we can see qualitatively that the time at which this behaviour is seen is
roughly $\propto {1\over g^2\mu}$.

Thus, 
even though we clearly see non--perturbative behaviour at low $k_t$, the 
apparent saturation of these modes at large times has no impact on the
behaviour of the energy density at late times. As $\mu$ 
is decreased, the magnitudes of the energy density appear to converge to 
a fixed value. This behavior of the energy has a purely kinematic reason:
for any fixed finite rapidity the proper time $\tau$ asymptotically approaches
the real time $t$, with respect to which the energy is conserved.
LPTh also predicts this behavior.

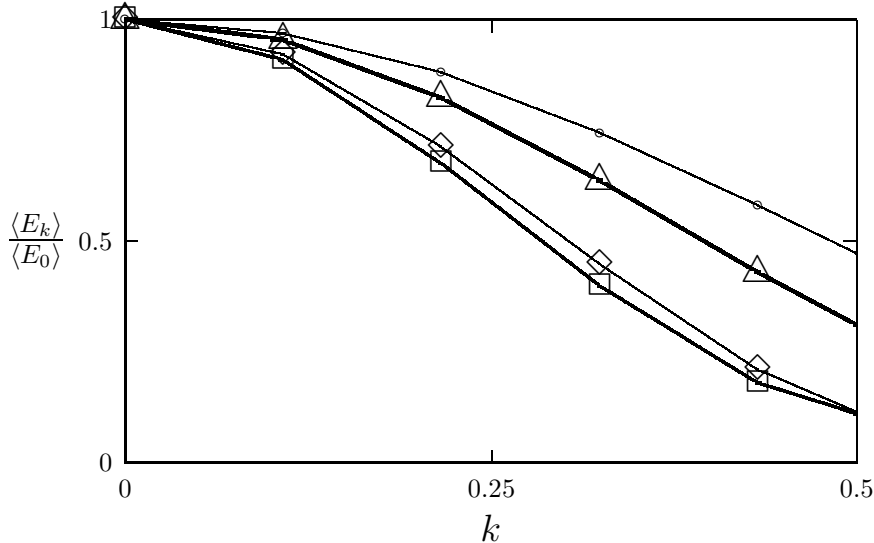
\begin{figure}[ht]
\setlength{\unitlength}{0.240900pt}
\ifx\plotpoint\undefined\newsavebox{\plotpoint}\fi
\sbox{\plotpoint}{\rule[-0.200pt]{0.400pt}{0.400pt}}%
\begin{picture}(1350,900)(0,0)
\begin{Large}
\font\gnuplot=cmr10 at 10pt
\gnuplot
\sbox{\plotpoint}{\rule[-0.200pt]{0.400pt}{0.400pt}}%
\put(181.0,163.0){\rule[-0.200pt]{4.818pt}{0.400pt}}
\put(161,163){\makebox(0,0)[r]{0}}
\put(1310.0,163.0){\rule[-0.200pt]{4.818pt}{0.400pt}}
\put(181.0,511.0){\rule[-0.200pt]{4.818pt}{0.400pt}}
\put(161,511){\makebox(0,0)[r]{0.5}}
\put(1310.0,511.0){\rule[-0.200pt]{4.818pt}{0.400pt}}
\put(181.0,859.0){\rule[-0.200pt]{4.818pt}{0.400pt}}
\put(161,859){\makebox(0,0)[r]{1}}
\put(1310.0,859.0){\rule[-0.200pt]{4.818pt}{0.400pt}}
\put(181.0,163.0){\rule[-0.200pt]{0.400pt}{4.818pt}}
\put(181,122){\makebox(0,0){0}}
\put(181.0,839.0){\rule[-0.200pt]{0.400pt}{4.818pt}}
\put(756.0,163.0){\rule[-0.200pt]{0.400pt}{4.818pt}}
\put(756,122){\makebox(0,0){0.25}}
\put(756.0,839.0){\rule[-0.200pt]{0.400pt}{4.818pt}}
\put(1330.0,163.0){\rule[-0.200pt]{0.400pt}{4.818pt}}
\put(1330,122){\makebox(0,0){0.5}}
\put(1330.0,839.0){\rule[-0.200pt]{0.400pt}{4.818pt}}
\put(181.0,163.0){\rule[-0.200pt]{276.794pt}{0.400pt}}
\put(1330.0,163.0){\rule[-0.200pt]{0.400pt}{167.666pt}}
\put(181.0,859.0){\rule[-0.200pt]{276.794pt}{0.400pt}}
\put(41,511){\makebox(0,0){\Large${{\langle E_k\rangle}\over{\langle E_0\rangle}}$}}
\put(755,61){\makebox(0,0){\Large$k$}}
\put(181.0,163.0){\rule[-0.200pt]{0.400pt}{167.666pt}}
\put(181,859){\usebox{\plotpoint}}
\multiput(181.00,857.92)(2.265,-0.499){107}{\rule{1.904pt}{0.120pt}}
\multiput(181.00,858.17)(244.049,-55.000){2}{\rule{0.952pt}{0.400pt}}
\multiput(429.00,802.92)(0.856,-0.499){287}{\rule{0.784pt}{0.120pt}}
\multiput(429.00,803.17)(246.372,-145.000){2}{\rule{0.392pt}{0.400pt}}
\multiput(677.00,657.92)(0.677,-0.500){365}{\rule{0.641pt}{0.120pt}}
\multiput(677.00,658.17)(247.669,-184.000){2}{\rule{0.321pt}{0.400pt}}
\multiput(926.00,473.92)(0.747,-0.500){329}{\rule{0.698pt}{0.120pt}}
\multiput(926.00,474.17)(246.552,-166.000){2}{\rule{0.349pt}{0.400pt}}
\multiput(1174.00,307.92)(1.167,-0.499){131}{\rule{1.031pt}{0.120pt}}
\multiput(1174.00,308.17)(153.859,-67.000){2}{\rule{0.516pt}{0.400pt}}
\put(181,859){\raisebox{-.8pt}{\makebox(0,0){$\Diamond$}}}
\put(429,804){\raisebox{-.8pt}{\makebox(0,0){$\Diamond$}}}
\put(677,659){\raisebox{-.8pt}{\makebox(0,0){$\Diamond$}}}
\put(926,475){\raisebox{-.8pt}{\makebox(0,0){$\Diamond$}}}
\put(1174,309){\raisebox{-.8pt}{\makebox(0,0){$\Diamond$}}}
\sbox{\plotpoint}{\rule[-0.400pt]{0.800pt}{0.800pt}}%
\put(181,859){\usebox{\plotpoint}}
\multiput(181.00,857.09)(1.953,-0.501){121}{\rule{3.300pt}{0.121pt}}
\multiput(181.00,857.34)(241.151,-64.000){2}{\rule{1.650pt}{0.800pt}}
\multiput(429.00,793.09)(0.766,-0.501){317}{\rule{1.425pt}{0.121pt}}
\multiput(429.00,793.34)(245.043,-162.000){2}{\rule{0.712pt}{0.800pt}}
\multiput(677.00,631.09)(0.645,-0.500){379}{\rule{1.232pt}{0.121pt}}
\multiput(677.00,631.34)(246.443,-193.000){2}{\rule{0.616pt}{0.800pt}}
\multiput(926.00,438.09)(0.817,-0.501){297}{\rule{1.505pt}{0.121pt}}
\multiput(926.00,438.34)(244.876,-152.000){2}{\rule{0.753pt}{0.800pt}}
\multiput(1174.00,286.09)(1.607,-0.502){91}{\rule{2.747pt}{0.121pt}}
\multiput(1174.00,286.34)(150.299,-49.000){2}{\rule{1.373pt}{0.800pt}}
\put(181,859){\raisebox{-.8pt}{\makebox(0,0){$\Box$}}}
\put(429,795){\raisebox{-.8pt}{\makebox(0,0){$\Box$}}}
\put(677,633){\raisebox{-.8pt}{\makebox(0,0){$\Box$}}}
\put(926,440){\raisebox{-.8pt}{\makebox(0,0){$\Box$}}}
\put(1174,288){\raisebox{-.8pt}{\makebox(0,0){$\Box$}}}
\sbox{\plotpoint}{\rule[-0.500pt]{1.000pt}{1.000pt}}%
\sbox{\plotpoint}{\rule[-0.600pt]{1.200pt}{1.200pt}}%
\put(181,859){\usebox{\plotpoint}}
\multiput(181.00,856.26)(3.927,-0.500){54}{\rule{9.600pt}{0.121pt}}
\multiput(181.00,856.51)(228.075,-32.000){2}{\rule{4.800pt}{1.200pt}}
\multiput(429.00,824.26)(1.364,-0.500){172}{\rule{3.570pt}{0.120pt}}
\multiput(429.00,824.51)(240.590,-91.000){2}{\rule{1.785pt}{1.200pt}}
\multiput(677.00,733.26)(0.957,-0.500){250}{\rule{2.598pt}{0.120pt}}
\multiput(677.00,733.51)(243.607,-130.000){2}{\rule{1.299pt}{1.200pt}}
\multiput(926.00,603.26)(0.860,-0.500){278}{\rule{2.367pt}{0.120pt}}
\multiput(926.00,603.51)(243.088,-144.000){2}{\rule{1.183pt}{1.200pt}}
\multiput(1174.00,459.26)(0.938,-0.500){156}{\rule{2.555pt}{0.120pt}}
\multiput(1174.00,459.51)(150.696,-83.000){2}{\rule{1.278pt}{1.200pt}}
\put(181,859){\makebox(0,0){$\triangle$}}
\put(429,827){\makebox(0,0){$\triangle$}}
\put(677,736){\makebox(0,0){$\triangle$}}
\put(926,606){\makebox(0,0){$\triangle$}}
\put(1174,462){\makebox(0,0){$\triangle$}}
\sbox{\plotpoint}{\rule[-0.500pt]{1.000pt}{1.000pt}}%
\sbox{\plotpoint}{\rule[-0.200pt]{0.400pt}{0.400pt}}%
\put(181,859){\usebox{\plotpoint}}
\multiput(181.00,857.92)(5.722,-0.496){41}{\rule{4.609pt}{0.120pt}}
\multiput(181.00,858.17)(238.434,-22.000){2}{\rule{2.305pt}{0.400pt}}
\multiput(429.00,835.92)(2.041,-0.499){119}{\rule{1.726pt}{0.120pt}}
\multiput(429.00,836.17)(244.417,-61.000){2}{\rule{0.863pt}{0.400pt}}
\multiput(677.00,774.92)(1.313,-0.499){187}{\rule{1.148pt}{0.120pt}}
\multiput(677.00,775.17)(246.616,-95.000){2}{\rule{0.574pt}{0.400pt}}
\multiput(926.00,679.92)(1.089,-0.499){225}{\rule{0.970pt}{0.120pt}}
\multiput(926.00,680.17)(245.986,-114.000){2}{\rule{0.485pt}{0.400pt}}
\multiput(1174.00,565.92)(1.028,-0.499){149}{\rule{0.921pt}{0.120pt}}
\multiput(1174.00,566.17)(154.088,-76.000){2}{\rule{0.461pt}{0.400pt}}
\put(181,859){\circle{12}}
\put(429,837){\circle{12}}
\put(677,776){\circle{12}}
\put(926,681){\circle{12}}
\put(1174,567){\circle{12}}
\end{Large}
\end{picture}
\vskip -0.6cm
\caption{Energy per mode k normalized to energy of zeroth mode as a function 
of k in GeV units. From bottom to top, the curves correspond to proper times
of $\tau=0,1.4,2.8,5.6$ fermis. Here $\mu=0.41$ GeV. }
\label{ecorrvsk}
\end{figure}

We now describe preliminary estimates of the energy distribution, obtained
from energy-energy correlators, as described in Section 4.
The computation was performed for
an $80\times 80$ lattice and for $\mu = 0.41$ GeV. With the small data sample
available (50 independent configurations for the values of proper time $\tau$
considered), we found that the correlation functions $C(x_t)$ are consistent 
with 0 at distances $|x_t|\geq 12$ in lattice units. We therefore
approximated $C(x_t)$ as 0 for $|x_t|\geq 12$. While the quality of our data  
at this point does not permit a quantitative description of the gluon
distribution, the data do help us address the question as to whether the
unusual behavior of the soft modes, as shown in 
Figure~\ref{a2kvsphtminkl160}, has observable consequences.

In Fig.~\ref{ecorrvstau}, we plot
the time dependence of the ratio of the energy of
the kth mode to the total energy of the system (measured directly). 
The ratio, for the momenta
considered ($0, 108$ and $432$) MeV, goes to a constant at late times. 
Since Fig.~\ref{ehistl160} clearly shows that the energy dies off as
$1/\tau$, our result suggests that at late times the energy of the
low $k_t$ modes must die off as $1/\tau$ too.  We therefore have an indication
that the apparent saturation of
the field intensities in Fig.~\ref{a2kvsphtminkl160} is not meaningful
and is a gauge artifact.

In Figure~\ref{ecorrvsk}, we plot the total energy per mode normalized to the 
energy 
of the zeroth mode as a function of $k$ for the proper times (from bottom to 
top) $\tau=0,1.4,2.8,5.6$ fm.
Interestingly, the ratio plotted gets
flatter at larger values of $\tau$. We interpret this to mean that the share 
of the energy in the lower-$k_t$ modes decreases at late times. This is
another indication of a benign behavior of soft modes at late times,
contrary to what is suggested by Figure~\ref{a2kvsphtminkl160}.

\section{Summary and outlook}

In this paper, we have performed a non-perturbative study of the 
production and space--time evolution of gluon mini--jets in the 
central region of nuclear
collisions at very high energies. This program was proposed 
in Ref.~\cite{RajKrasnitz} and a brief discussion of our results can be
found in Ref.~\cite{RajKrasnitz2}. 

Our work is based on a classical effective field theory description of
the small $x$ modes in nuclei at very high energies. This effective
theory contains a scale $\mu$ which is proportional to the large gluon
density at small $x$. The large gluon density ensures that even if the
coupling is weak, the fields may be highly non--perturbative.  Since
the approach is classical, it is possible to study the real time
evolution of these non--perturbative modes in a nuclear collision. The
approach has the attractive feature that it may eventually provide a
self-consistent picture of high energy nuclear collisions both before
and after the nuclei collide.

At large transverse momenta, $k_t\gg \alpha_S\mu$, the predictions of the
effective theory for gluon production should agree with perturbative 
mini-jet calculations. As discussed in Section 5, our lattice 
results agree with the lattice perturbation theory analogue of the 
continuum mini-jet predictions. At smaller transverse momenta, 
(in particular for large $\mu$) we see significant deviations from 
perturbation theory. We also notice that the field intensities, 
$|A(k_t,\tau)|^2$, of the small $k_t$ modes do not die off at the late times
studied but appear to saturate.

It is difficult to interpret the behaviour of the small $k_t$ modes since 
they are ``off-shell'': their dispersion relation $\omega(k_t)$ is 
non--trivial. Unlike the large momentum limit, the field intensities for
the small $k_t$ modes does not have the simple interpretation of being the 
number distribution of produced gluons.  As discussed in the previous section 
it is therefore important to look at gauge invariant quantities which 
can be interpreted straightforwardly. In the previous section we discussed
a number of gauge invariant equal time spatial correlators. These 
correspond to correlators of components of
the energy-momentum tensor. In particular, we suggested a gauge invariant
estimate of the energy distribution on the lattice as a function of
time. 
In principle, we should be able to compute the range of energy-energy 
corelators measured in high energy collisions (see Basham et
al.~\cite{Brownetal} and references therein). A more extensive discussion of
these correlators will be presented at a later date~\cite{RajKrasnitz3}.

An important issue not addressed in this work is that of equilibration--
do the gluons produced equilibrate? We saw in Fig.~\ref{ehistl160} that the
energy density shows the expected behaviour after times $\propto 1/g^2\mu$.
This however is no proof of thermalization. In the classical approach, 
clearly the high $k_t$ modes are not ``thermal''. Whether this is the case
for the softer modes needs to be studied further. It would be interesting 
to relate this approach to that of M\"uller and collaborators~\cite{Muller,
MullRev,BasMullPos}.

We would like to stress that this work is not a quantitative study of
gluon production in heavy ion collisions but a qualitative study of 
non-perturbative effects that may be important in the central region of
these collisions. Changes that can be made in future to bring our results
closer to experiment include a) changing the gauge group to SU(3), b) 
relaxing the boost invariance assumption, and c) modifying the boundary 
conditions. Regarding the last point, we have used periodic boundary 
conditions. Modifying these would, for instance, be useful in computing the
Poynting flux through the nuclear surface.

Though our choice of gauge $A^\tau=0$ is convenient from the point of 
view of a lattice simulation, results in this gauge do not lend themselves
easily to a diagrammatic interpretation. In this regard it might be useful 
also to consider a similar computation in 
Coulomb gauge where this is the case~\cite{Yuri}. 
Results of such computation will be presented at a later date.

\section*{Acknowledgments}

We would like to thank the Universidade do Algarve (RV) and the
Niels Bohr Institute (AK) for
their kind hospitality. We would also like to thank M. Gyulassy, 
Yuri Kovchegov, Alex Kovner, B. M\"uller, Rob Pisarski, J. Randrup, 
Dirk Rischke and X.-N. Wang for useful comments and discussions. 
We would like to thank Larry McLerran for his critical
input and encouragement during the course of this work.
RV's work is supported by 
the Danish Research Council and the Niels Bohr Institute. Both AK and 
RV  acknowledge support provided by the Portuguese Funda\c c\~ao para a 
Ci\^encia e a Technologia,
grants CERN/S/FAE/1111/96 and CERN/P/FAE/1177/97.

\section*{Appendix A: Numerical method}
\vspace*{0.3cm}

In integrating equations of motion of a classical gauge theory it is important
to ensure that the redundant gauge degrees of freedom do not become observable 
through
numerical errors. Hence the integration scheme must respect the Gauss
constraints of the theory. To this end, the algorithm of choice for the
problem at hand is the leapfrog algorithm~\cite{CG}. This integration
scheme requires that the Hamiltonian be a sum of gauge-invariant kinetic $K$
(dependent on fields only) and potential $V$ (dependent on conjugate momenta
only) terms, and under this condition, has the advantage of respecting the
Gauss
constraints exactly~\cite{thalgs}. Our lattice Hamiltonian (\ref{hl})
obviously has the necessary $K+V$ form, thus the leapfrog algorithm is
applicable.

The light-cone Hamiltonian (\ref{hl}) depends explicitly on the proper time
$\tau$. For this reason, the standard leapfrog scheme, designed for
time-independent Hamiltonian functions, requires some minor adaptation in the
present case. In order to explain the suitable version of the algorithm, we
collectively denote the $E$, $p$ momentum\footnote{Strictly speaking, $E$
variables are not conjugate momenta. Indeed, their Poisson brackets with each
other do not necessarily vanish. However, the scheme as presented only requires
that $\{K,E\}=0$, and $E$ need not be momentum variables.} variables by $P$ and
the $U$, $\Phi$ field variables by $Q$, respectively. The $K$ and $V$ terms in
the Hamiltonian then have the form $K(P,\tau)$ and $V(Q,\tau)$. In the
following, the argument $\tau$ of $K$ or $V$ is thought of as a {\it fixed
parameter}. With this notation, the leapfrog step of size $\Delta$, propagating
the system in proper time from $\tau$ to $\tau+\Delta$, has the following form.
\begin{enumerate}
\item Integrate the equations ${\dot Q}=\{K(P,\tau),Q\}$ between $\tau$ and
$\tau+\Delta/2$.
\item Integrate the equations ${\dot P}=\{V(Q,\tau+\Delta/2),P\}$ between
$\tau$ and $\tau+\Delta$.
\item Integrate the equations ${\dot Q}=\{K(P,\tau+\Delta),Q\}$ between
$\tau+\Delta/2$ and $\tau+\Delta$.
\end{enumerate}
Here the integration of the equations of motion is assumed to be performed
exactly at every substep. It can be readily seen that the step as described
is exactly reversible, {\it i.e.}, beginning with the final values of $P,Q$
and performing the step with $\Delta$ replaced by $(-\Delta)$, one arrives at
the initial $P,Q$. This property of the leapfrog algorithm, evident for a
time-independent Hamiltonian, is preserved here by suitably choosing the
explicit proper-time argument of $K$ and $V$ at each substep. The time
reversibity guarantees that the integration error, obviously of the order
no lower than $\Delta^2$, is in fact ${\cal O}(\Delta^3)$.

In order to impose the initial conditions for the proper-time Hamiltonian
evolution, we first need to solve the lattice Poisson equation (\ref{latpoi}).
We do so using the overrelaxation method \cite{recipes}. The consistency of
Eq.~\ref{latpoi} requires that the zero-momentum component of the color charge
density $\rho$ vanish. We therefore first generate the color charge 
distribution as a normal deviate (\ref{distriblat}), then subtract from every
$\rho_{q,j}$ the spatial average $\sum_j\rho_{q,j}/N$.

We fix the Coulomb gauge on initial configurations (see also 
the discussion in section 2.2). The lattice Coulomb gauge reads
\be
{\rm Tr} \left[\left\{ \sum_n \left(U_{j,\nhat}^\prime
-{U_{j,\nhat}^\prime}^\dagger\right) -\sum_n\left(U_{j-n,\nhat}^\prime-
{U_{j-n,\nhat}^\prime}^\dagger\right)\right\}\sigma^a\right] =0 \, ,
\label{lattcoul}
\ee
where $U_{j,\nhat}^\prime$ is the link matrix which satisfies the Coulomb
gauge condition (\ref{lattcoul}). We use the standard 
overrelaxation method \cite{mog} for gauge fixing.

\section*{Appendix B: Lattice perturbation theory}
\vspace*{0.3cm}

At large transverse momenta, a test of our lattice results is that they agree
with those of lattice perturbation theory. Here we present a derivation of
classical gluon production in lattice perturbation theory. In the continuum
limit, we will show that our result agrees with Eq.~(27) of
Ref.~\cite{KLW}. We also obtain an expression for the initial kinetic energy
on the lattice. We will later compare these analytical expressions to the
full lattice results at large transverse momenta.

We restrict our discussion to the gauge group SU(2) and consider the
initial condition (\ref{ucondp}) for the link matrix $U$
\be
U=(U^{(1)}+U^{(2)})({U^{(1)}}^\dagger+{U^{(2)}}^\dagger)^{-1} \, .\nonumber
\ee
Here the labels of the nuclei are written as superscripts. Now recall that
since $U_{j,n}^{1,2}$ are  pure gauges, they can be written as (see 
Eq. \ref{pgs})
\be
U_{j,n}^{(i)} = V_j^{(i)}{V_{j+n}^{(i)}}^\dagger \, ,\nonumber
\ee
where $V_j^{(i)}=\exp(i\Lambda_j^{(i)})$ and $i= 1,2$.  Hence, 
\be
U_{j,n}^{(i)} = 1 + i \left(\Lambda_j^{(i)}-\Lambda_{j+n}^{(i)}\right)
+\left(\Lambda_j^{(i)}\Lambda_{j+n}^{(i)}-{1\over 2}\left[
(\Lambda_j^{(i)})^2 + (\Lambda_{j+n}^{(i)})^2\right]\right) + 
O(\Lambda^3) \, .
\label{uexp}
\ee
The pure gauge solution of the Yang--Mills equations for a single nucleus
dictates that $\nabla_{\perp}^2 \Lambda^{(i)} = \rho^{(i)}$ (see also 
Eq.~\ref{puresoln}). Therefore $\Lambda\sim O(\mu)$ and we have kept terms in
the expansion above up to $O(\mu^2)$. 
Then substituting Eq.~\ref{uexp} in the expression for $U$, we obtain
\be
U_{j,n} = I + i L_{j,n} + {1\over 2}
\left(Q_{j,n}-Q_{j,n}^\dagger-L_{j,n}^2\right) + O(\mu^3) \, ,
\label{Uexpand}
\ee
where $I$ is the SU(2) identity matrix,
\be
L_{j,n} = \sum_{i=1}^{2} \left(\Lambda_j^{(i)}-\Lambda_{j+n}^{(i)}\right)
\equiv \alpha_{j,\nhat}^{(1)} + \alpha_{j,\nhat}^{(2)} \, ,
\label{Lform}
\ee
and 
\be
Q_{j,n} =\sum_{i=1}^2 \left[\Lambda_j^{(i)}\Lambda_{j+n}^{(i)} -
{1\over 2} \left((\Lambda_j^{(i)})^2 + (\Lambda_{j+n}^{(i)})^2\right)\right]
\, .
\ee

We work in the $A_\tau=0$ gauge, supplemented by the Coulomb gauge condition
(\ref{lattcoul}) at $\tau=0$. The Coulomb gauge-transformed link matrix 
$U^\prime$ is related to the original link $U$ as
\be
{U_{j,n}}^\prime = W_j U_{j,n} W_{j+n}^\dagger \, . \nonumber
\ee
This is analogous to the transformation carried out in the continuum
perturbation theory derivation discussed in section 2.2.
For $\mu=0$ the Coulomb gauge is satisfied by the trivial configuration $U=I$, 
Hence the gauge transformation $W_j$ can be expanded in powers of small $\mu$:
$$W_j = \exp\left(I + i\mu \xi_j + i\mu^2\eta_j+{\cal O}(\mu^3)\right).$$
We now need
to determine $\xi$ and $\eta$ from the Coulomb gauge condition (\ref{lattcoul}).
Writing out ${U}^\prime$ to order $\mu^2$, we obtain
\be
{U_{j,n}}^\prime &=& I + i\mu\left(\xi_j-\xi_{j+n}+L_{j,n}\right)
+\mu^2 \Bigg( i(\eta_j-\eta_{j+n})-{1\over 2} (\xi_j^2 + \xi_{j+n}^2)
\nonumber \\
&+&\xi_j\xi_{j+n} + (L_{j,n}\xi_{j+n}-\xi_j\L_{j,n})+
{1\over 2} (Q_{j,n}-Q_{j,n}^\dagger-L_{j,n}^2)\Bigg)\, .
\label{lincoul}\ee
To lowest order, the Coulomb gauge condition (\ref{lincoul}) implies that
\be
\xi_j-\xi_{j+n} = - L_{j,n} \longrightarrow \xi_j =
-\left(\Lambda_j^{(1)}+\Lambda_j^{(2)} \right) \, .
\ee
Now consider the Coulomb gauge condition to order $\mu^2$. We need to
determine $\eta_j$ to obtain $W_j$ to that order. After some algebra, one
can show that
\be 
\eta_j = {1\over {2i}}{1\over \Delta}\left\{[\Lambda_j^{(1)},\Lambda_{j-n}^
{(2)}+\Lambda_{j+n}^{(2)}] + (1)\leftrightarrow (2)\right\} \, .
\ee
Above, $\Delta$ is the usual lattice Laplacian
\be
\Delta(l) = 2\sum_{n=1,2} (1-\cos(l_n)) \, , \nonumber
\ee
and $l_n = k_n \cdot a$. 
Having computed $\xi$ and $\eta$ using the Coulomb gauge fixing condition, 
we are now in a position to compute $\alpha^\prime$--the lattice analog of
the field $\epsilon^i$ in section 2. We have
\be
2i\,\alpha_{j,n}^\prime = {U_{j,n}}^\prime - 
{{U_{j,n}}^\prime}^\dagger 
\equiv 2i \left(\eta_j - \eta_{j+n}\right) -  \left(
[\Lambda_j^{(1)},\Lambda_{j+n}^{(2)}] + (1)\leftrightarrow (2)\right)
 + {\cal O}(\mu^3) \, .
\ee

To determine the field intensity we need to compute the lattice Fourier 
transform of the above expression. Using the Fourier transform of 
$\Lambda$,
\be
\Lambda_j^{(1),b} = {1\over N^2}\sum_{p=-(N-1)/2}^{(N-1)/2} \exp(2\pi i
\vec{p}\cdot \vec{x_j}/L) {\tilde{\Lambda}}_l^{(1),b}
\, , 
\ee
after quite some algebra, we obtain the result that
\be
{\alpha^\prime}_{l,n}^a &=& {\epsilon^{abc}\over {N^2}} 
\Bigg[\sum_{\l^\prime}\Bigg\{ 
\left({(1-e^{il_n})\over \Delta(l)}\right)\, 
\left(\sum_{n^\prime} (-)2\sin\left({l_{n^\prime}
\over 2}\right)\sin\left({l_{n^\prime}\over 2}-l_{n^\prime}^\prime\right)
\right)\nonumber \\
&-& ie^{il_{n}/2} \sin\left({l_n\over
2}-l_n^\prime\right)\Bigg\}
 {\tilde{\Lambda}}_{l^\prime}^{(1),b}{\tilde{\Lambda}}_{l-l^\prime}^
{(2),c}\Bigg] \, .
\label{discr}
\ee
It is useful to compare this result to the corresponding continuum expression.
Here and in the following the continuum result is obtained by setting
$\alpha_l^\prime \rightarrow
\alpha_k^\prime a/g$, $l_n\rightarrow k_i a$ and $\sum_{l^\prime}\rightarrow
L^2\int {d^2 k^\prime \over {(2\pi)^2}}$ and letting $a\rightarrow 0$.
In the case of Eq. \ref{discr} this prescription gives
\be
{\alpha^\prime}_{k,n}^a =\epsilon^{abc} \int {d^2 k^\prime\over{(2\pi)^2}} 
\left( {k_n k_{n'}\over k^2}-\delta_{ij}\right){\tilde\Lambda}_
{k^\prime}^{(1),b} i(k_{n'}-k_{n'}^\prime){\tilde \Lambda}_{k-
k^\prime}^{(2),c} \, ,
\ee
This result is identical to the Fourier transform of Eq.~(27) in 
Kovner, McLerran and Weigert~\cite{KLW}.

We can now compute the field intensity directly. It is defined as
\be
|\alpha_l^\prime|^2 = \sum_n \langle{\alpha^\prime}_{l,n}^a 
{\alpha^\prime}_{-l,n}^a\rangle_\rho \, ,
\label{dscr2}
\ee
where $\langle\cdots\rangle_\rho$ denotes the Gaussian averaging over the 
sources. On the lattice, the continuum condition 
$\nabla_\perp^2 \Lambda^{(i)} = \rho^{(i)}$ translates into
\be
{\tilde \Lambda}_l^{(i)} = -\mu L {\eta_l^{(i)}\over \Delta(l)} \, ,
\label{lamfour}
\ee
where $\eta_l = {\tilde \rho}_l a^2/\mu L$. Hence 
$\langle\eta_l\rangle_\rho = 0$ 
and $\langle\eta_l\eta_{l^\prime}\rangle_\rho = \delta_{l,-l^\prime}$.
Substituting Eq.~\ref{discr} in Eq.~\ref{dscr2} and performing the averaging
over $\rho$'s, we obtain finally the following expression for the field 
intensity
\begin{equation}
|\alpha_l^\prime|^2 = {{3\mu^4}\over{2\Delta(l)}}{\sum_{l'}}^\prime
{{\Delta(2l'-l)\Delta(l)-\left[\Delta(l'-l)-\Delta(l')\right]^2}
\over{\Delta^2(l')\Delta^2(l'-l)}}\, .
\label{dcrfin}
\end{equation}
Here ${\sum_{l'}}^\prime$ means that terms with $l'=l$ or $l'=0$ are omitted.
Taking the continuum limit of Eq.~\ref{dcrfin}, just as we did for 
Eq.~\ref{discr} and keeping in mind that
$\mu$ in this expression is given in units of $1/ag^2$ ({\it cf}. Section 4), 
one recovers the 
corresponding continuum expression of Ref.~\cite{KLW} for $N_c=2$.
In the section on our numerical results
we compare our lattice results for the
field intensity to Eq.~\ref{dcrfin}.

We now compute the initial kinetic energy of the
scalar field $\Phi$ in lattice perturbation theory. This can also be checked
against our lattice results in weak coupling and provides yet another test of
the numerics. In the continuum, using Eq.~\ref{dinitial} and 
Eq.~\ref{initial}, we have 
\be
p^a p^a =\epsilon^{abc} \epsilon^{ade} \alpha_{\perp i}^{(1),b} 
\alpha_{\perp i}^{(2),c}\alpha_{\perp j}^{(1),d}\alpha_{\perp j}^{(1),e} \, ,
\ee
with $i=1,2; j=1,2$. On the lattice, writing 
\be
\alpha_{\perp i}^{(1),b} \alpha_{\perp i}^{(1),c} \longrightarrow 
\sum_n \left\{ (\alpha_\perp^{(1),b} \alpha_\perp^{(2),c})_{j,\nhat} + 
(\alpha_\perp^{(1),b}\alpha_\perp^{(2),c})_{j-n,\nhat} \right\} \, ,
\ee 
and using $\alpha_{j,\nhat}^{(i)} = (\Lambda^{(i)}(x_j+a\nhat)-
\Lambda^{(i)} (x_j))/a$
\be
\alpha_{j,\nhat} = -{\mu\over {N}} \sum_{\vec{k}} {\left(\exp(il_n)-1\right)
  \eta_l\over {\Delta(l)}} e^{2\pi i \vec{k}\cdot {\vec{x}}_j} \, ,
\ee
we have (after Gaussian averaging over $\eta$'s)
\be
p^a p^a = 6\left(\mu\over {N}\right)^4 \sum_{n,n^\prime} 
\left[ \left(\sum_{\vec{k}} {\sin(l_n)\sin(l_{n^\prime})\over \Delta^2 (l)}
\right)^2 + 16\left(\sum_{\vec{k}} {\sin^2({l_n\over 2})\sin^2({l_{n^\prime}
\over 2})\over \Delta^2 (l)}\right)^2 \right] \, .\nonumber\\
\label{dike}
\ee
Let us take a look at the continuum limit of this expression. For $n\neq
n^\prime$, the first term vanishes. For $n=n^\prime$, this term is $\propto 
\mu^4 \log^2 (L/a)$; it is logarithmically divergent in the infrared. The
second term is infrared safe for all $n,n^\prime$. Hence
\be
p^a p^a \longrightarrow A + B \log^2 (L/a) \, ,
\ee 
where $A$ and $B$ are constants which may be determined from Eq.~\ref{dike}. 

Thus far we have not specified precisely what the lattice expansion parameter
is. In the continuum, perturbation theory is applicable when $\alpha_S\mu/k_t
\ll 1$. In Ref.~\cite{RajGavai} the lattice expansion parameter for a single
nucleus was estimated numerically to be $g^2\mu L$. That this is also the
case here can been seen from Eqs.~\ref{uexp} and \ref{lamfour}.

Finally, the reader may have noted that our lattice perturbation theory
results were computed at $\tau=0$. As noted in section 2, in weak coupling
the spatial and temporal distributions factorize. The spatial distributions 
of the high momentum modes are therefore completely specified at $\tau=0$.

\end{document}